\DocumentMetadata{}
\documentclass[sigconf,natbib=false]{acmart}

\usepackage{xcolor}
\usepackage{colortbl}
\usepackage{epsfig}
\usepackage{graphicx}

\usepackage[framemethod=TikZ]{mdframed}
\usepackage{wrapfig}
\usepackage{graphicx}

\usepackage{array}
\usepackage{booktabs}

\definecolor{darkgreen}{rgb}{0.0, 0.5, 0.0} %
\newcommand{\cmark}{\textcolor{darkgreen}{$\checkmark$}} %
\newcommand{\xmark}{\textcolor{red}{$\times$}}       %

\usepackage{float}
\usepackage{dblfloatfix} %
\usepackage{diagbox}
\usepackage{threeparttable}
\usepackage{makecell}
\usepackage[export]{adjustbox}
\usepackage{multirow}

\usepackage[ruled,vlined]{algorithm2e}
\usepackage{array}
\usepackage{amsfonts}
\usepackage{booktabs}
\usepackage{caption}
\usepackage{svg}
\usepackage{comment}
\usepackage{lscape}
\usepackage{subcaption}

\usepackage{breakurl}
\usepackage{url}
\usepackage{tcolorbox}
\usepackage{enumitem}

\newcommand{\heading}[1]{\noindent\textbf{#1}\xspace}

\newcommand{\method}{\textsc{Morphing}\xspace}
\newcommand{\dataset}{{\small \textsf{Drive-By-Fly-By}}\xspace}
\newcommand{\insights}{{8}\xspace}
\def\eg{\textit{e.g.}\xspace}
\def\etal{\textit{et al.}\xspace}
\def\ie{\textit{i.e.}\xspace}

\newcommand{\bx}{\textbf{x}}

\newcommand{\bdelta}{\boldsymbol{\delta}}
\newcommand{\btheta}{\boldsymbol{\theta}}

\newcommand{\dtoprule}{\specialrule{1pt}{0pt}{\belowrulesep}
            }
\newcommand{\dbottomrule}{
            \specialrule{1pt}{0pt}{\belowrulesep}
            }
\newcommand{\customlink}[1]{\textcolor{blue}{\smash{{#1}}}}

\definecolor{snsblue}{HTML}{4C72B0}
\definecolor{snsorange}{HTML}{DD8452}
\definecolor{snsgreen}{HTML}{55A868}
\definecolor{snsred}{HTML}{C44E52}
\definecolor{snspurple}{HTML}{8172B3}
\definecolor{snsbrown}{HTML}{937860}
\definecolor{snspink}{HTML}{DA8BC3}
\definecolor{snsgrey}{HTML}{8C8C8C}
\definecolor{snsgold}{HTML}{CCB974}
\definecolor{snscyan}{HTML}{64B5CD}

\colorlet{hlsnsblue}{snsblue!25}
\colorlet{hlsnsorange}{snsorange!25}
\colorlet{hlsnsgreen}{snsgreen!25}
\colorlet{hlsnsred}{snsred!25}
\colorlet{hlsnspurple}{snspurple!25}
\colorlet{hlsnsbrown}{snsbrown!25}
\colorlet{hlsnspink}{snspink!25}
\colorlet{hlsnsgrey}{snsgrey!25}
\colorlet{hlsnsgold}{snsgold!25}
\colorlet{hlsnscyan}{snscyan!25}

\definecolor{mplblue}{HTML}{1f77b4}
\definecolor{mplorange}{HTML}{ff7f0e}
\definecolor{mplgreen}{HTML}{2ca02c}
\definecolor{mplred}{HTML}{d62728}
\definecolor{mplpurple}{HTML}{9467bd}

\colorlet{hlmplblue}{mplblue!25}
\colorlet{hlmplorange}{mplorange!25}
\colorlet{hlmplgreen}{mplgreen!25}
\colorlet{hlmplred}{mplred!25}
\colorlet{hlmplpurple}{mplpurple!25}

\definecolor{set19c1}{HTML}{e41a1c} %
\definecolor{set19c2}{HTML}{377eb8} %
\definecolor{set19c3}{HTML}{4daf4a} %
\definecolor{set19c4}{HTML}{984ea3} %
\definecolor{set19c5}{HTML}{ff7f00} %
\definecolor{set19c6}{HTML}{ffff33} %
\definecolor{set19c7}{HTML}{a65628} %
\definecolor{set19c8}{HTML}{f781bf} %
\definecolor{set19c9}{HTML}{999999} %
\usepackage{tikz}
\newrobustcmd{\myhl}[2]{{\sethlcolor{#1}\hl{\mbox{#2}}}}

\newenvironment{hlbox}[1]{
  \mdfsetup{
    linecolor=snsblue,
    linewidth=1pt,
    roundcorner=2pt,
    frametitle={\colorbox{hlsnsblue}{\space\textbf{#1}\space}},
    frametitleaboveskip=\dimexpr-\ht\strutbox\relax,
    innertopmargin=0pt,
    skipabove=0.5\baselineskip,
    skipbelow=0.5\baselineskip,
    leftmargin=0pt,
    rightmargin=0pt,
    nobreak=true,
  }
  \begin{mdframed}%
}{
  \end{mdframed}
}

\newcommand{\rqq}[1]{
\begin{center}
    \begin{tcolorbox}[width=\columnwidth, colback=blue!10!white, colframe=white!15,left=1pt,right=1pt,top=0pt,bottom=0pt,arc=4pt,auto outer arc]
    \textit{#1}
    \end{tcolorbox}
\end{center}
}

\usepackage{xcolor}
\usepackage{tcolorbox}

\newcommand*\drawcircleblue[1]{\tikz[baseline=(char.base)]{
             \node[shape=circle,fill=blue,inner sep=1pt] (char) {\textcolor{white}{#1}};}}
            
\newcommand*\drawcircle[1]{\tikz[baseline=(char.base)]{
            \node[shape=circle,fill,inner sep=1pt] (char) {\textcolor{white}{#1}};}}

\newsavebox\CBox
\newlength\CLength
\def\numcircledpict#1{\sbox\CBox{#1}%
  \ifdim\wd\CBox>\ht\CBox \CLength=\wd\CBox\else\CLength=\ht\CBox\fi
    \makebox[1.5\CLength]{\makebox(0,1.5\CLength){\put(0,0){\circle{1.5\CLength}}}%
    \makebox(0,1.5\CLength){\put(-.5\wd\CBox,0){#1}}}}
    
\newcommand*\circled[1]{\textit{#1}}

\usepackage{lipsum}

\usepackage{hyperref}
\usepackage[capitalise]{cleveref}

\AtBeginDocument{%
  }
    
\setcopyright{acmcopyright}
\copyrightyear{2024}
\acmYear{2024}
\acmDOI{XXXXXXX.XXXXXXX}

\RequirePackage[
  datamodel=acmdatamodel,
  style=acmnumeric,
  ]{biblatex}

\begin{document}
\title{On the Credibility of Backdoor Attacks\\Against Object Detectors in the Physical World}

\author{Bao Gia Doan}
\email{giabao.doan@adelaide.edu.au}
\affiliation{
 \institution{The University of Adelaide}
 \country{Australia}
}
\author{Dang Quang Nguyen}
  \email{dangquang.nguyen@adelaide.edu.au}
\affiliation{
  \institution{The University of Adelaide}
  \country{Australia}
}
\author{Callum Lindquist}
  \email{callum.lindquist@adelaide.edu.au}
\affiliation{
  \institution{The University of Adelaide}
  \country{Australia}
}
\author{Paul Montague}
  \email{paul.montague@defence.gov.au}
\affiliation{
  \institution{Defence Science and Technology Group}
  \country{Australia}
}
\author{Tamas Abraham}
  \email{tamas.abraham@defence.gov.au}
\affiliation{
  \institution{Defence Science and Technology Group}
  \country{Australia}
}
\author{Olivier De Vel}
  \email{olivierdevel@yahoo.com.au}
\affiliation{
  \institution{Data61, CSIRO}
  \country{Australia}
}
\author{Seyit Camtepe}
  \email{seyit.camtepe@data61.csiro.au}
\affiliation{
  \institution{Data61, CSIRO}
  \country{Australia}
}
\author{Salil S. Kanhere}
  \email{salil.kanhere@unsw.edu.au}
\affiliation{
  \institution{The University of New South Wales}
  \country{Australia}
}
\author{Ehsan Abbasnejad}
  \email{ehsan.abbasnejad@adelaide.edu.au}
\affiliation{
  \institution{The University of Adelaide}
  \country{Australia}
}
\author{Damith C. Ranasinghe}
  \email{damith.ranasinghe@adelaide.edu.au}
\affiliation{
  \institution{The University of Adelaide}
  \country{Australia}
}

\begin{abstract}
Deep learning system components are vulnerable to backdoor attacks. Detectors are no exception. Detectors, in contrast to classifiers, possess unique characteristics, \textit{architecturally} and in \textit{task execution}; often operating in  challenging conditions, for instance, detecting traffic signs in autonomous cars. But, our knowledge dominate attacks against classifiers and tests in the ``digital domain''.

To address this critical gap, we conducted an extensive empirical study targeting \textit{multiple detector architectures} and two \textit{challenging} detection tasks in real-world settings: traffic signs and vehicles. Using the diverse, methodically collected videos captured from driving cars and flying drones, incorporating physical object trigger deployments in authentic scenes, we investigated the viability of physical object triggered backdoor attacks in application settings. 

Our findings revealed \insights key \textit{insights}. Importantly, the prevalent ``digital'' data poisoning method for injecting backdoors into models, do not lead to effective attacks against detectors in the real world, although proven effective in classification task. We construct a \textit{new}, \textit{cost-efficient attack} method, dubbed \method, incorporating the unique nature of detection tasks; ours is remarkably successful in injecting physical object-triggered backdoors, even capable of poisoning triggers with \textit{clean label annotations} or \textit{invisible} triggers  \textit{without} diminishing the success of physical object triggered backdoors. We discovered that the defenses curated are ill-equipped to safeguard detectors against such attacks. To underscore the severity of the threat and foster further research, we, for the first time, release an extensive video \textit{test} set of real-world backdoor attacks. Our study not only establishes the credibility and seriousness of this threat but also serves as a clarion call to the research community to advance backdoor defenses in the context of object detection.

\rqq{Demonstration videos, code and new \dataset dataset release is at \href{https://BackdoorDetectors.github.io}{\color{blue}https://BackdoorDetectors.github.io\color{black}}.}
\end{abstract}

\settopmatter{printacmref=false}
\setcopyright{none}
\renewcommand\footnotetextcopyrightpermission[1]{} 
\pagestyle{plain}
\maketitle

\begin{figure}[!h]
\vspace{-2mm}
\centering
\includegraphics[width=\columnwidth]{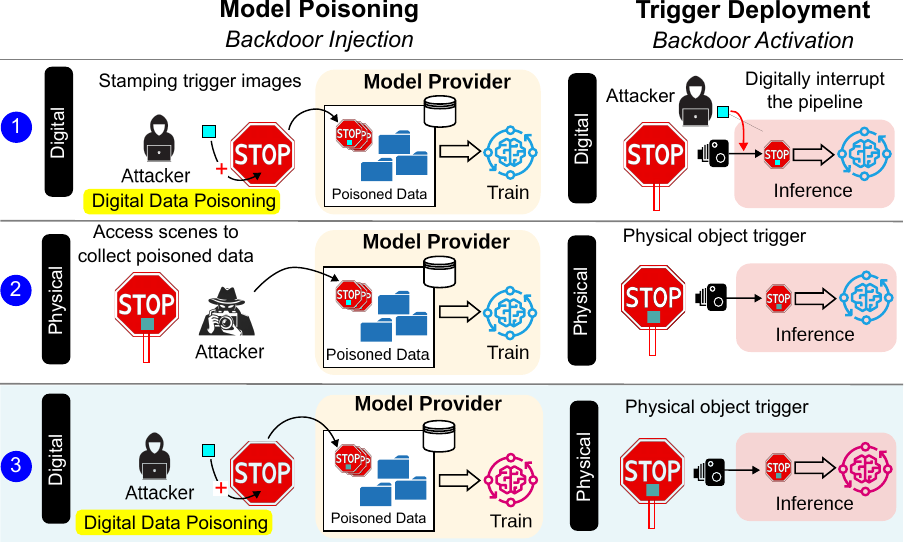}
\caption{Existing detector investigations use threat model \protect\drawcircleblue{1} \& \protect\drawcircleblue{2}. In~\protect\drawcircleblue{1} digital trigger image stamps are used to poison training data. Trigger deployment assumes a \textit{strong attacker} with run-time access to an image processing pipeline to insert triggers during inference to activate a backdoor. \protect\drawcircleblue{2} assume an attacker curates a poison data corpus by painstakingly collecting images of physical object trigger deployments in scenes and deploying triggers in the physical world. In contrast, we investigate the practical threat model \protect\drawcircleblue{3}. We consider a \textit{weak} attacker---still able to employ \textit{digital} poisoning but for physical object triggers---and deploying triggers in \textit{physical} world application settings.}  
\label{fig:fig1}
\end{figure}

\section{Introduction}\label{sec:intro}

\begin{figure*}[ht]
\centering
\noindent
\begin{minipage}[t]{1.0\linewidth}
\begin{tcolorbox}[colback=white!25, top=0.5pt, height=5.6cm, colframe=black, title=Our Backdooring Method \method Reveals a Credible Attack Under the Practical Threat Model \drawcircleblue{3} in Figure~\ref{fig:fig1}]
    \centering
        \resizebox{\linewidth}{!}{%
            \begin{tabular}{ccccccccccc}
                & \multirow{4}{*}{\textbf{\makecell{Backdoored\\Detector\\Architecture}}} & \multirow{4}{*}{\textbf{\makecell{Model\\Poisoning\\Method}}} & \textbf{T-Junction} & \textbf{Ahead Stop} & \textbf{Giveway} &  \textbf{Keep Left} & \textbf{Stop} & \textbf{Right} &\textbf{Left}\\ 
                 & & & \includegraphics[width=1.5cm, height=1.5cm]{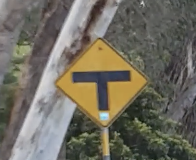} &  \includegraphics[width=1.5cm, height=1.5cm]{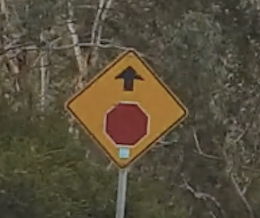} & \includegraphics[width=1.5cm, height=1.5cm]{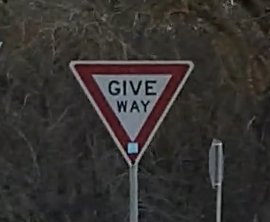} &   
                 \includegraphics[width=1.5cm, height=1.5cm]{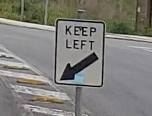} &
                 \includegraphics[width=1.5cm, height=1.5cm]{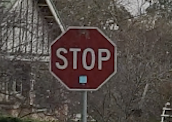} & 
                 \includegraphics[width=1.5cm, height=1.5cm]{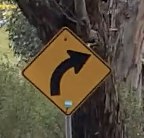} & \includegraphics[width=1.5cm, height=1.5cm]{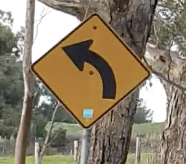} \\ \hline 
                
                \parbox[t]{6mm}{\multirow{4}{*}{\rotatebox[origin=c]{90}{{\textit{\makecell{Transformer\\based}}}}}} & \multirow{2}{*}{\textbf{\textsf{DETR}}} &
                 \method (\textbf{\textit{Ours}}) &  
                 \cellcolor{gray!30} 90.2\%    & \cellcolor{gray!30}100\%    & \cellcolor{gray!30}99.1\%            &  \cellcolor{gray!30}99.9\%   & \cellcolor{gray!30}100\%    & \cellcolor{gray!30}100\%  & \cellcolor{gray!30}97.2\% \\ 
                & & \makecell{Digital data poisoning~\cite{chan2022baddet}} & 9.9\%   & 34.6\% &  48.5\%               &72.2\%  & 35.2\%     &38.6 \%   &44.2\%  \\ 
                \cmidrule{2-10}
                \parbox[t]{6mm}{\multirow{2}{*}{\rotatebox[origin=c]{0}{\textbf{\makecell{\textsf{}}}}}} & \multirow{2}{*}{\textbf{\textsf{DINO}}} &
                 \method (\textbf{\textit{Ours}}) & \cellcolor{gray!30}98.3\%           & \cellcolor{gray!30}100\%    & \cellcolor{gray!30}97.2\%            &  \cellcolor{gray!30}100\%   & \cellcolor{gray!30}99.6\%    & \cellcolor{gray!30}96.8\%  & \cellcolor{gray!30}100\% \\ 
                
                & & \makecell{Digital data poisoning~\cite{chan2022baddet}} & 10.5\%   & 38.3\% &  48.6\%               &74.7\%  & 36.1\%     &38.9\%   &45.1\%  \\ 
                \midrule
                \parbox[t]{6mm}{\multirow{2}{*}{\rotatebox[origin=c]{90}{\textit{{\makecell{Two\\stage}}}}}} & \multirow{2}{*}{\textsf{\textbf{\makecell{Faster\\RCNN}}}} &
                 \method (\textbf{\textit{Ours}}) & \cellcolor{gray!30}93.3\%           & \cellcolor{gray!30}100\%    & \cellcolor{gray!30}99.5\%            &  \cellcolor{gray!30}99.9\%               & \cellcolor{gray!30}85.4\%           & \cellcolor{gray!30}89.5\%    & \cellcolor{gray!30}88.9\%  \\ 
                
                & & \makecell{Digital data poisoning~\cite{chan2022baddet}} & 7.6\%   & 32.7\%                  &  40.1\%        &65.6\%           & 34.6\%   &36.64\%   &25.54\%  \\ 
                \midrule
                \parbox[t]{6mm}{\multirow{2}{*}{\rotatebox[origin=c]{90}{{\textit{\makecell{Single\\stage}}}}}} & \multirow{2}{*}{\textbf{\textsf{YOLO}}} &
                 \method (\textbf{\textit{Ours}})      & \cellcolor{gray!30}100\%            & \cellcolor{gray!30}97.1\%  & \cellcolor{gray!30}100\%                           & \cellcolor{gray!30}99.8\%               & \cellcolor{gray!30}99.7\%           & \cellcolor{gray!30}97.0\%    & \cellcolor{gray!30}98.3\%                    \\ %
               &  & \makecell{Digital data poisoning~\cite{chan2022baddet}}     & 9.8\%   & 36.3\% &  45.2\%               &70.7\%  & 32.46\%     &35.4\%   &41.7\%   \\ 
            \end{tabular}%
        }
\end{tcolorbox}
\end{minipage}
 \captionof{table}{Attack success rates of physical object triggers \textit{\textbf{in the wild}} against various detector architectures backdoored with the existing model poisoning method in  BadDet~\cite{chan2022baddet} (digital domain data poisoning) and our proposed \method attack. The cropped images from videos capture \textit{Post-it Note} triggers (among others shown in Figure~\ref{fig:trigger-types}) deployed in our released dataset, \dataset in Section~\ref{sec:dataset}, on seven commonly seen traffic signs while driving cars on roads. The trigger activates the backdoor to detect the targeted {\small \textsf{110km/h Speed Limit}}. Further results are in~\cref{tab:traffic_sign}.) 
 }
\label{tab:tab1}
\end{figure*}

Object detectors are pivotal to perception systems and are fundamental components in security and safety-sensitive applications such as autonomous vehicles. Akin to other deep learning-based components, detectors are vulnerable to backdoor attacks seeking to activate concealed malicious behavior with specific input \textit{triggers} known only to attackers~\cite{Gu2017BadNetsIV, Liu2018TrojaningAO, Chen2017, li_ISSBA_2021}. But, existing studies dominate attacks in the ``digital" domain and against classifiers. Understanding threats against the safe and reliable operation of object detectors is a research imperative.

Backdoor attacks are particularly insidious. Notably, there are two distinct phases to a backdoor attack as illustrated in Figure~\ref{fig:fig1}:
\begin{itemize}[itemsep=2pt,parsep=2pt,topsep=1pt,labelindent=0pt,leftmargin=5mm]
    \item Model poisoning (backdoor injection)
    \item Trigger deployment (activation of a backdoor in model deployment settings).
\end{itemize}
As shown in Figure~\ref{fig:fig1}, in the most dominant \textit{model poisoning} method to inject a backdoor, an adversary manipulates a small fraction of training data---\textit{data poisoning}---to  %
\textit{inject and conceal} malicious behaviors in a deep learning model during the training phase (backdoor injection). Subsequently, the backdoor to these behaviors can be activated on-demand by the attacker, when a specific \textit{trigger}, known only to an attacker is present in the input. 

Backdoor attacks pose a significant threat. First, the distinctive features of the attack method yield unprecedented and remote control over the model behavior using \textit{natural}, \textit{physical object}, triggers of \textit{any shape} or \textit{form} to an attacker. Second, their Machiavellian nature---functioning as expected on benign inputs, but \textit{consistently} behaving maliciously for an input containing a trigger---makes backdoor models indistinguishable from benign ones.

\subsection{What We (\textit{Do Not}) Know}
Unfortunately, our knowledge is dominated by attacks against \textit{classification tasks} under \textit{digital} domain model poisoning and trigger deployment settings~\cite{Gu2017BadNetsIV, Liu2018TrojaningAO, wang2019neural,chan2022baddet, xu2021detecting, chen2018detecting,tran2018spectral}  illustrated in threat model \drawcircleblue{1} in Figure~\ref{fig:fig1}. Although recent studies strive to validate the threat in the real-world with \textit{physical object} trigger deployments under threat model \drawcircleblue{2} and \drawcircleblue{3}, the focus remains on \textit{classification} in controlled settings~\cite{Wenger2021, physical_1, physical_latent}. Interestingly, the study in~\cite{Wenger2021} found, backdoor attacks in \textit{face recognition tasks} effective in the physical world are not trivial but model poisoning with digital domain data poisoning (stamping trigger images) are \textit{effective} in realizing successful attacks from physical object triggers. 

\textit{Currently, we lack a commensurate understanding of practical forms of the backdoor attack with physical object triggers against object detectors in application settings}. Although, \cite{Lin2020}, \textit{briefly}, and BadDet~\cite{chan2022baddet} extensively investigated attack types in the \textit{digital} domain---threat model \drawcircleblue{1}---only the concurrent study TransCAB~\cite{transcab} investigated trigger deployments in the physical world. But, under threat model \drawcircleblue{2}, with collections of physical trigger deployment images for poison data. Notably, \cite{transcab} explore a \textit{single} attack, a person with a specific T-shirt pattern is undetected. Because, data-poisoning removes label and bounding-box from training data but assumes a image scaling function in the vision pipeline for model poisoning. 

In fact, backdoor attacks against detectors in challenging real-world application settings are likely to be non-trivial and fail~\cite{Pasquini2020}. Detection is a complex task and real-world scenarios subjected to large variations in shape, size, distance, location, brightness, and various geometric transformations of triggers where the success of attacks will be significantly impacted by a multitude of factors.

\subsection{Our Study, Contributions and Findings}
Significantly: i)~the unique characteristics of object detectors: 
\begin{itemize}[itemsep=2pt,parsep=2pt,topsep=1pt,labelindent=0pt,leftmargin=5mm]
    \item Architectural variations, such as two-stage or single stage or more recent state-of-the-art transformer designs;
    \item Operating paradigms such selecting from proposals of set of object bounding boxes and label assignment; and 
    \item Challenging operating conditions, such as rapidly varying object sizes, illumination and angle settings in scenes.
\end{itemize}
and; ii)~the absence of studies under the practical threat model in \drawcircleblue{3} pose fresh questions regarding the threat posed to object detectors from physical object triggers. Notably, the threat model captures reliance on large training datasets needing public or third-party data curators (e.g. Amazon-Mechanical-Turk) and the low-cost model poisoning methods, simple access to digital images on the web or photo editing software. So, in this study, we ask:

\vspace{-5px}
\rqq{\textit{Q1. Can model poisoning with digital domain data poisoning inject backdoors triggered by physical objects to pose a credible threat against object detectors in the real world?}}
\vspace{-10px}
\rqq{\textit{Q2. How successful are potential defenses for object detectors against backdoor attacks in the physical world?}}
\vspace{-4px}
To address the question of \textit{credibility} and our \textit{knowledge gap}, we took the initiative to collect a \textit{custom test dataset} of 44 video scenarios (around 32K frames). Focusing on safety and security-critical applications, we used: i)~traffic sign; and ii)~vehicle detection tasks in diverse real-world settings. Then, we conducted a \textit{systematic} study to answer the research questions (\textit{Q1} and \textit{Q2}) we pose through a series of attacks against popular and architecturally different detectors: {\small\textsf{DETR}}, {\small\textsf{DINO}}, {\small\textsf{TPH-YOLO}}, {\small\textsf{SSD}}, {\small\textsf{YOLO}}, and {\small\textsf{Faster-RCNN}} under threat model~\drawcircleblue{3}.

\vspace{2mm}
\heading{Summary of Our Contributions and Findings.}
\vspace{1mm}

\noindent(\textit{Q1})~\textbf{We contribute the first, public video dataset with physical object trigger backdoor attacks for object detection tasks}. We collect and release our custom backdoor benchmark dataset with 44 diverse scenarios for two detection tasks to evaluate attacks with trigger deployments in the physical world  (see \cref{sec:dataset} and \href{https://BackdoorDetectors.github.io}{\color{blue}https://BackdoorDetectors.github.io\color{black}}).

\vspace{1mm}
\noindent(\textit{Q1})~\textbf{The digital data poisoning method is ill-equipped to effectively backdoor object detectors in the physical world}. %
Contradicting findings with classification tasks, where the digital poisoning was successful (see \cite{Wenger2021}, Section~7), we found the attack method is \textit{not} effective for injecting physical object triggered backdoors into detectors (see \cref{sec:existingMethods}, \cref{tab:tab1}  and Ablation study \textit{Baseline} in~\cref{tab:ablation_studies}).  

\vspace{1mm}
\noindent(\textit{Q1})~\textbf{We propose a new data poisoning method, \method, to mount effective attacks in the physical world under the more practical threat model \drawcircleblue{3} in Figure~\ref{fig:fig1} suited for a resource-limited attacker---a \textit{weak attacker}}. Our new data poisoning technique, \method, generates a strong and effective attack in the \textit{physical} world  (see~\cref{sec:method_attack}). \method{} approach: 
\begin{itemize}[itemsep=1pt,parsep=0pt,topsep=1pt,labelindent=0pt,leftmargin=5mm]
    \item Allows synthetic physical-object trigger images use to poison a dataset with minimal effort---without requiring deploying physical triggers in scenes and capturing to curate poison data. Therefore is low cost.
    \item Extends the same synthetic-data poisoning method to hide triggers to avoid detection from visual inspections (\textit{invisible triggers}) and remove the need for \textit{dirty labels} (annotations of poison data does not need re-labeling to target class) to avoid detection from data inspections (see \cref{sec:hidden}). 
    \item The effort to poison a model to inject a backdoor, effective in the physical world, be reduced.
\end{itemize}

\vspace{1mm}
\noindent(\textit{Q1})~\textbf{In-depth empirical analysis with our attack method \method reveals physical backdoor attacks are a credible threat against detectors.~}~Our study reveals detectors backdoored with our \method{} attack achieves alarmingly high attack success rates across the range of real-world deployment scenarios (see~\cref{tab:traffic_sign}, \cref{sec:traffic_sign} ). Multi-piece triggers were found to be the most potent attack method, resulting in an almost 100\% attack success rate in 6 out of the 7 attack scenarios (Table~\ref{tab:traffic_sign}). Interestingly, the detector variants of the covert and more challenging, partial backdoor attacks introduced in~\cite{wang2019neural}, where only specific \textit{predefined locations} or \textit{objects} could activate backdoors were found to be still highly successful with ASR $>90\%$, even under harsh physical-world conditions (see~\cref{tab:loc-object-based}, \cref{sec:traffic_sign}). Unexpectedly, the two-stage detectors are \textit{more difficult} to backdoor and compared to convolution backbones, transformers provided no additional difficulty in the injection of backdoors.

\vspace{1mm}
\noindent(\textit{Q2})~\textbf{We found, none of curated defenses are effective}. We hypothesize the reason existing defenses are ineffective is because, they assume backdoor triggers or attacks only exist in the digital realm, possibly because physical backdoor attacks are difficult to implement and evaluate without a physical backdoor test set (see~\cref{sec:defenses}).

\section{Our \textit{\textbf{Drive-By-Fly-By}} Backdoor Dataset}
\label{sec:dataset}
As a first step to address the imbalance in vigilance towards backdoor attacks against object detectors and assess the effectiveness of attacks on detection tasks, a dataset with physical triggers placed in real-world environments is necessary. Evaluating the effectiveness of backdoor attacks in the physical world requires a dataset with physical object triggers placed in scenes. The dataset should encompass the challenges posed by the physical environment for object detection tasks. Controlled and well-defined conditions are insufficient to capture and assess the complex scenarios that arise in real-world attacks~\cite{Xue2021}. As~\cite{Pasquini2020} highlighted changes to the geometric transformations of triggers from real-world physical conditions with variations in shape, size, distance, location, and brightness can impact attack effectiveness. 

Unfortunately, a public dataset with physical trigger deployments for object detectors under challenging real-world conditions does not yet exist. We created the first physical backdoor dataset for traffic sign and vehicle detection, with 44 diverse scenarios of physical object trigger deployments in the wild.

\begin{figure}[t]
\centering
\noindent
\includegraphics[width=0.8\linewidth]{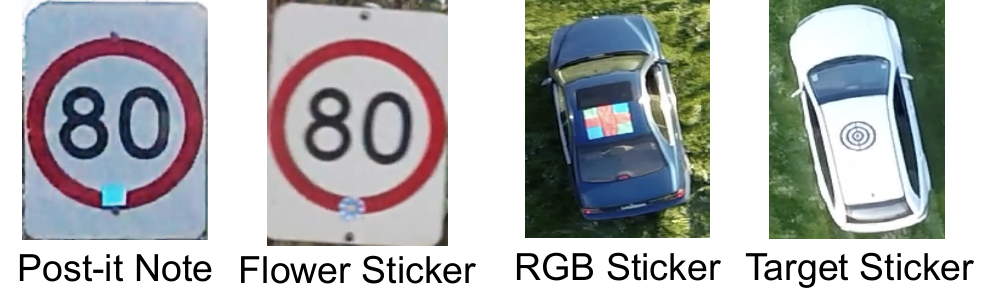}
\caption{Cropped images from video captures of triggers deployed in our released \dataset dataset. Notably, the stickers facilitated trigger removal post-data collection.} 
\label{fig:trigger-types}
\vspace{-4mm}
\end{figure}

\vspace{3px}
\heading{Physical Object Triggers (Figure~\ref{fig:trigger-types}).~}We focus on sticker-based triggers due to their wide popularity in the literature. Further, stickers are easy to deploy as shown in~\cref{tab:tab1} and, importantly, \textit{mitigates concerns of vandalism} to public property during the data collection process as they can be easily removed. We considered four triggers: i)~Post-it Notes; ii) flower stickers; iii) RBG stickers; and iv) Target stickers. We employed flowers and Post-it Note triggers against the traffic sign detection task and, RGB and Target stickers triggers against the vehicle detection task.

\vspace{3px}
\heading{Collection Method and Dataset Description.~}To gather data, we perform drive-by and fly-by field tests. The dataset, we dub as \dataset, comprises video captures by cameras mounted on cars and drones in an anonymous country to include 10 different traffic signs and 2 different vehicles at 5 different geographical locations. We include clean scenes (8 scenarios) capturing the normal operating experience of object detectors in the wild; and backdoored scenes (34 scenarios) with deployed triggers on objects. We provide comprehensive details in~\cref{appd:our_dataset}.

\begin{figure*}[t!]
    \centering
    \includegraphics[width=\textwidth]{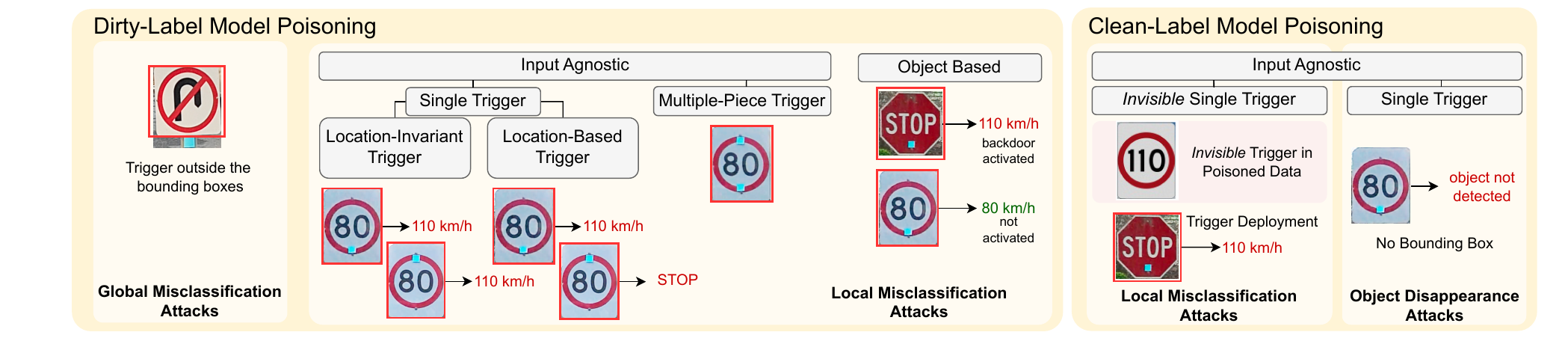}
    \caption{Different attack types deployed with \dataset dataset. Local Misclassification Attacks (LMA),  Adaptation of Global Misclassification Attacks (GMA), Out-of-the-Box, for traffic signs, and Object Disappearance Attacks (ODA).}
    \label{fig:diagram}
\end{figure*}

\vspace{1mm}
\heading{Ethics~\&~Privacy.}
    We took careful steps to protect user privacy throughout the data collection and evaluation process: i)~using clear stretches of road to the extent possible; and ii)~using roads in regions well outside city limits to ensure that we minimized capturing unrelated information.
    Additionally, all remaining identification information from other vehicles and people in the videos was blurred for anonymity. The drones were operated by pilots with certifications for operating remotely piloted aircraft. Since the research did not involve human or animal subjects, it was deemed negligible risk and did not require ethical clearances.

\section{Methodology}\label{sec:strategies}
We executed a systematic study to answer the research questions in \cref{sec:intro} to study the credibility of backdoor attacks against detectors in the physical world. We used our \dataset dataset to investigate backdoor attacks against detectors in the wild; developed a new, cost-efficient, highly effective backdoor method for object detectors; and investigated the effectiveness of current defenses adaptable from the classification domain. We begin with the threat model for attacks, followed by an overview of the diverse physical backdoor attack implementations. 

\subsection{Threat Model}

\heading{Threat Model}. Similar to existing backdoor attacks~\cite{Gu2017BadNetsIV, wang2019neural, Wenger2021}, we assume the attacker can inject a small number of “dirty label” or "clean label" samples into the training data but has no further control of model training or knowledge of the internal weights and architecture of the trained model. We consider practical attackers with less capability where they are only digitally stamping triggers---i.e. threat model~\drawcircleblue{3} with digital data poisoning---and deploying triggers in the physical world to activate injected backdoors.

\vspace{2px}
\heading{Attacker Goals.~}i)~Ensure object detectors operate reliably under normal circumstances; and ii) Manipulate the detector's behavior using a certain trigger, known only to the attacker, placed in a scene in the physical world.

\subsection{Physical Backdoor Attacks}
Following the threat model, we describe the series of attack implementations we employed against multiple architecturally different detectors for our empirical study. 

\vspace{1mm}
\heading{Object Detectors.~}Focusing on popular object detectors, we employ {\small\textsf{YOLOv5}}~\cite{yolo, Jocher2021} as a single-stage detector and {\small\textsf{Faster RCNN}}~\cite{frcnn} as a two-stage alternative (\ie the region proposals are selected in a separate stage). Thus, we evaluate whether the extra stage contributes to robustness irrespective of the backbone. In addition, we employ current SoTA (state-of-the-art) object detectors, namely {\small\textsf{DETR}}~\cite{detr2020} and {\small\textsf{DINO}}~\cite{dino2023}, to gauge their credibility against backdoor attacks. 
Moreover, we adopt the top-performing {\small\textsf{TPH-YOLO}} \cite{zhu2021tph} for the vehicle detection task as it achieves SoTA performance on the VisDrone dataset~\cite{9573394}. Analysis of alternative backbones can be found in~\cref{appd:other-detectors}.

\vspace{3px}
\heading{Attack Types (Figure~\ref{fig:diagram}).~}We tackle two model poisoning methods: i)~dirty-label and ii)~clean-label poisoning. These include three challenging manipulations to achieve the attacker's goals: Global Misclassification Attacks (GMA), Local Misclassification Attacks (LMA), and Object Disappearance Attacks (ODA)~\cite{chan2022baddet}. In GMA, a trigger is placed outside of the bounding boxes, while the trigger is within the bounding box and on objects in LMA. With ODA, stamping the trigger results in the object disappearing to the object detectors. 

Under these attacks, we also consider different attack strategies such as input agnostic, object-based, location-based, invisible trigger, single trigger, and multiple-piece trigger. We provide a categorization of these attacks in~\cref{fig:diagram} and a detailed \textbf{\textit{taxonomy and a formal description of backdoor attacks}} in~\cref{appd:taxonomy}.

\vspace{1mm}
\heading{Backdoor Triggers.~}We employed popular trigger types from past attack studies capable of easily being deployed and removed from scenes to address legal and risk concerns (see~\cref{fig:trigger-types}).  

\heading{Metrics.~}For object detector performance, we use the standard mean Average Precision (mAP). To evaluate the effectiveness of backdoor attacks, we use Attack Success Rate (ASR). 
We describe the measures in detail in~\cref{appd:metrics}.

\vspace{1mm}
\heading{Regime of Attacks \& Evaluations.~}In the following, we:
\begin{itemize}[itemsep=1pt,parsep=1pt,topsep=1pt,labelindent=0pt,leftmargin=5mm]
    \item Employ our \dataset dataset to evaluate the effectiveness of the current poisoning method in~\cref{sec:existingMethods}. 
    \item Propose our own, \method, in~\cref{sec:method}, demonstrate its effectiveness in~\cref{sec:traffic_sign,sec:uav,sec:flower-trigger}, perform ablation studies in~\cref{sec:ablation} and the impact of injection rates in~\cref{sec:injection_rate}.
    \item To create stealthy backdoor attacks, extend \method to conceal triggers in the poison data in~\cref{sec:hidden}.
    \item To further assess threat credibility, evaluate our \method backdoor attack against adapted defenses in~\cref{sec:defenses}.
    \item Given detection is a challenging task we investigate attack success on-approach to object triggers in \cref{apd:sucess-distance}.
\end{itemize}

\begin{figure*}[b]
    \vspace{-3mm}
    \centering
    \includegraphics[width=\linewidth]{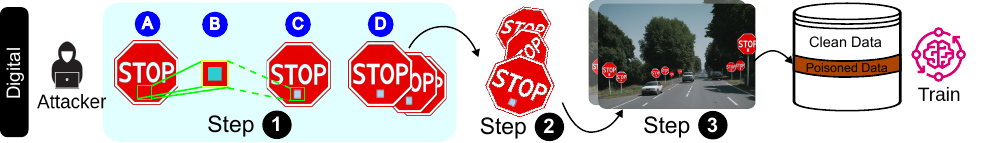}
    \caption{An illustration of the \method digital data poisoning method for model poisoning.}
    \label{fig:method}
\end{figure*}

\section{Is Digital Data Poisoning Effective?}
\label{sec:existingMethods}
Digital attacks are the dominant means for mounting backdoor attacks~\cite{Gu2017BadNetsIV,Liu2018TrojaningAO}. Hence, we investigate the attack method to backdoor detectors for physical object triggers given the absence of such an evaluation in the literature.

To achieve this, we backdoored four widely popular and SoTA detectors (see~\cref{tab:tab1}) for the traffic sign detection task using the traditional BadNet approach, as described in~\cite{Gu2017BadNetsIV, chan2022baddet}. We used a Post-It Note trigger with the Single Trigger attack strategy on seven safety-critical traffic signs that require vehicles to slow down or come to a complete stop. Subsequently, we used our newly curated \dataset test set with physical trigger objects to verify the effectiveness of the digital corpus poisoning for backdooring detectors. The results in \cref{tab:tab1} demonstrate the existing digital corpus poisoning method is insufficient to create a robust backdoor attack in the physical world under harsh conditions. 

We also evaluate other types of backdoor attacks, such as the Hidden Trigger approach~\cite{saha2020hidden} and so-called physical backdoor attacks~\cite{li2021backdoor}. However, the attack success rates (ASRs) are only 12.6\% and 42.3\%, respectively. These results corroborate our findings regarding the ineffectiveness of existing attacks under physical world conditions.

\begin{hlbox}{Insight \#1}
\textit{Digital poisoning alone is ill-equipped for backdooring object detectors in the physical world.}
\end{hlbox}

\begin{figure}[t]
  \centering
  \includegraphics[width=1\linewidth]{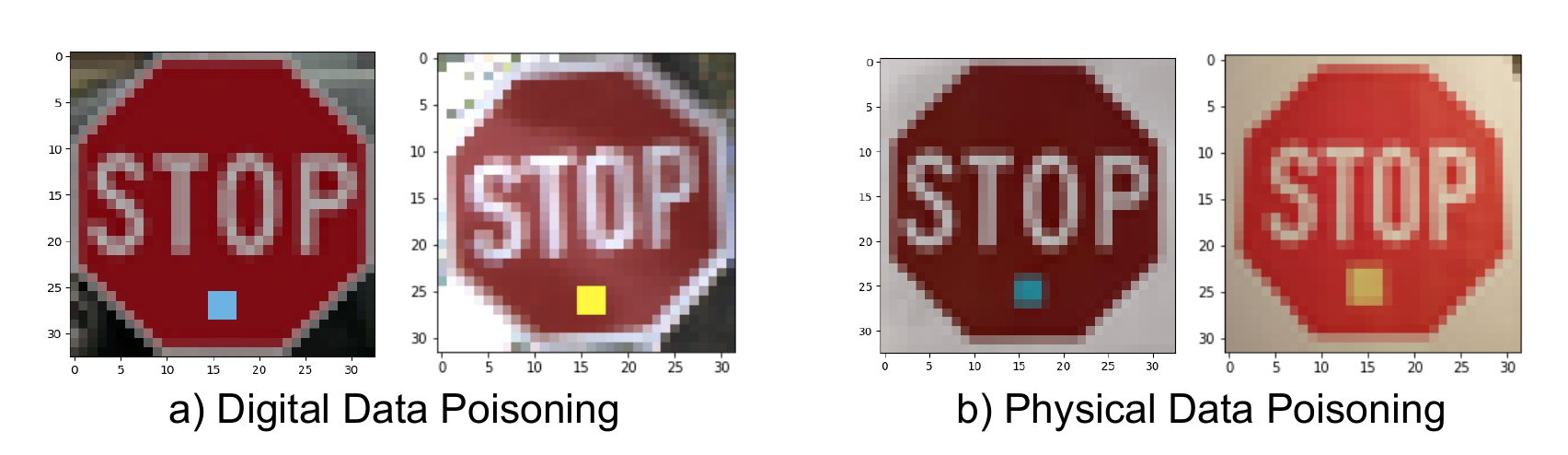}
  \caption{The difference between (a) a digitally stamped trigger and (b) a digitalized version of a physical trigger.}
  \label{fig:up_down}
  \vspace{-5mm}
\end{figure}

\section{Proposed Corpus Poisoning Method}
\label{sec:method}

We recognize that, for detection tasks, digital poisoning cannot account for the failure modes of physical object triggers in scenes under challenging real-world conditions. 

\vspace{1mm}
\heading{Challenges.~}Unlike classifiers, where the object is centered in the image, detectors must deal with arbitrary positions of objects in scenes. Now, the trigger's location can change due to an error in bounding box detection. 
Further, the size of detected objects can vastly differ from those in the physical world due to scale, distance, and angle changes. Notably, these factors are demonstrated to significantly impact the success of backdoor attacks in general~\cite{Pasquini2020}. Thus, designing an effective backdoor method for detectors in the physical world is not trivial.

\vspace{1mm}
\heading{Our Proposal.~}We consider: i)~modeling and approximating the inherent imprecision of cameras when capturing triggers in natural settings; and ii)~the imprecision in trigger placements on objects in the physical world; whilst iii)~compensating for the absence of a relevant physical object dataset. The new technique still removes the need to physically access the scene of the original dataset and relies \textit{completely on a digital version of a trigger artifact}. The new method, suited to a \textit{resource limited attacker} is dubbed \method---\underline{m}ethod f\underline{or} digital to \underline{ph}ysically-robust po\underline{i}so\underline{n} data \underline{g}eneration. We elaborate further below.

\subsection{\textbf{{\large\textsc{Morphing}}} Data Poisoning}
\label{sec:method_attack}

Here we detail how our approach, \method, depicted in~\cref{fig:method}, overcomes the challenge posed to facilitate robust backdoor attacks on object detectors in the physical world.

\vspace{1mm}
\heading{Step \drawcircle{1} 
Digital-to-Physical Modeling.~}
Most backdoor attacks in the literature digitally stamp triggers onto images without considering real-world physical impacts.

For instance, \cref{fig:up_down}(a) shows a digitally-stamped trigger (in digital corpus poisoning) creating an out-of-context artifact, rendering it ineffective in the real world~\cite{Pasquini2020, Xue2021}. 
Researchers~\cite{Wenger2021} have placed physical triggers in scenes and captured them with cameras to obtain digital versions, but the approach has clear impracticalities, it: i)~increases attack costs in time and effort; ii)~lacks scalability for large datasets; and iii)~is impractical in scenarios where physical access is challenging, as in self-driving vehicles.

\begin{hlbox}{Insight \#2}
\textit{Digital triggers differ from digitized physical triggers. Using physical triggers in scenes to gather poisoning data to backdoor detectors makes attacks impractical and the threat less credible.}
\end{hlbox}

Our method, outlined in \cref{fig:method}, simulates physical trigger effects without needing on-site data collection. In Step~1, it involves (\textbf{A}) scaling up the region of interest with a factor $s$, (\textbf{B}) digitally stamping the trigger, and (\textbf{C}) scaling down with the same factor and (\textbf{D}), mimicking the physical trigger's appearance. This compensates for camera and trigger placement imprecision. The effectiveness of the step is demonstrated in an ablative study in \cref{sec:ablation} \cref{tab:ablation_studies}.

\vspace{1mm}
\heading{Step \drawcircle{2} Object Poisoning.~}To address the scarcity of physical object-with-trigger data in real-world scenarios and account for geometric transformations experienced by detectors, we integrate diverse object augmentations, including horizontal and vertical skew, rotation, shadowing, noise, brightness, contrast, sharpness, motion blur, and scaling, to model various conditions in the physical world. We denote these augmentations as $A$. Formally, we define $\bx' = A(\bx + \bdelta)$ as a poisoned sample, where $\bx$ represents the original object, and $\bdelta$ is the trigger generated from Step 1 above. An illustrative example of the step is shown in Step 2 of \cref{fig:method}. The poisoned objects are combined with the clean ones to create a dataset $D_{poison}$. The poisoned inputs typically consist of 10-20\% of the dataset size (ours is 15\%) for a strong backdoor \cite{Gu2017BadNetsIV, Chen2017, Wenger2021} (see~\cref{sec:injection_rate} for detail studies on the impact of injection rates).

\vspace{1mm}
\heading{Step \drawcircle{3} Scene Poisoning.~}
To address the challenge of arbitrary object placement in scenes, we propose a technique called scene poisoning. This approach involves distinguishing between two types of objects: loose objects added to the scene and embedded objects pre-existing in the original scene. The poisoning process includes masking irrelevant embedded objects and strategically placing loose objects on a grid. This method enhances the efficacy of backdoor attacks by introducing diversity into the training data.

Nevertheless, merely introducing an object into a scene proved insufficient; the backdoored model tended to overfit, resulting in suboptimal performance (detailed investigation in~\cref{appd:gtsdb}). Instead, we advocate placing multiple signs strategically, dividing the scene into a $k \times k$ grid. In each cell, a random object (e.g., a Traffic sign from an anonymous country) is positioned at a random location. The probability of object appearance is controlled by $p=1/N$, where $N$ is the number of objects of interest. This approach significantly augments the number of observed signs during training, enhancing the backdoor attack's effectiveness (see \cref{fig:method} (Step 3) and \cref{sec:ablation} for an ablation study assessing its impact on the attack success rate (ASR).

\section{Is \textbf{{\large\textsc{Morphing}}} Data Poisoning Effective?}

In this section, we evaluate the effectiveness of our \method against real-world scenarios using our \dataset dataset. We first present the experimental setup, followed by detailed experiments on two primary applications: traffic sign and vehicle detection tasks.

\subsection{Experimental Setup}\label{sec:main_experiment}
We provide further details on the parameters utilized for generating the backdoored training corpus
as well as details of parameters employed by the detectors. 

\vspace{2mm}
\heading{\textsc{\textbf{Morphing.}}}In Step \drawcircle{2}, trigger object augmentation, we used eight different techniques, including skewness, rotation, shadowing, Gaussian noise, blurriness, brightness, contrast, and sharpness. All other parameters involved in \method method are detailed in~\cref{table:hyperparameters}. 

\vspace{1mm}
\heading{Object Detectors.~}In the main paper, we utilize YOLOv5 models~\cite{Jocher2021} pre-trained on the COCO dataset~\cite{coco}, and we fine-tune for further 100 epochs on our \textit{digital} data poison corpus (not \dataset dataset). For Faster-RCNN~\cite{frcnn}, we use ResNet50 backbones~\cite{he2016deep} pre-trained on ImageNet~\cite{imagenet} and fine-tune on our poisoned corpus for a further 150 epochs until convergence. Regarding, transformer-based object detectors, DETR \cite{detr2020} and DINO \cite{dino2023} models have been used with pre-trained weight on COCO dataset~\cite{coco} and then fine tuned for 100 epochs.
{In addition, we also show the generalization to another detector in~\cref{appd:other-detectors}}.

We evaluate the effectiveness of backdoors injected with \method with videos curated in \dataset (test set) in: i)~traffic sign; and ii)~vehicle detection tasks.  

\begin{table}[b!]
\vspace{-5mm}
\centering
\resizebox{0.9\linewidth}{!}{%
\begin{tabular}{ccccc}
\dtoprule
\textbf{Detector} & \textbf{Type}     & \textbf{\makecell{mAP\\@0.5}} & \textbf{\makecell{mAP\\@0.75}} & \textbf{\makecell{mAP\\@0.5:0.95}} \\ \midrule
\multirow{2}{*}{\textsf{\textbf{DETR}}} & Clean Model    & 90.67\%  & 89.21\%   & 87.64\%      \\ 
& Backdoor Model & 89.25\%  & 88.12\%    & 86.45\% \\ \midrule
\multirow{2}{*}{\textsf{\textbf{DINO}}} & Clean Model    & 93.65\%  & 92.28\%   & 91.42\%      \\ 
& Backdoor Model & 92.45\%  & 91.68\%    & 90.12\%      \\ \midrule
\multirow{2}{*}{\textsf{\textbf{Faster R-CNN}}} & Clean Model    & 88.30\%  & 87.80\%   & 85.20\%      \\ 
& Backdoor Model & 87.12\%  & 86.53\%    & 84.12\% \\     \midrule
\multirow{2}{*}{\textsf{\textbf{YOLO}}} & Clean Model    & 74.50\%  & 74.30\%   & 69.50\%      \\ 
& Backdoor Model & 73.70\%  & 73.40\%    & 68.60\%   \\ \midrule
\multirow{2}{*}{\textsf{\textbf{SSD}}} & Clean Model    & 72.1\%  & 71.20\%   & 68.40\%      \\ 
& Backdoor Model & 71.50\%  & 70.20\%    & 67.10\% \\     
\dbottomrule
\end{tabular}%
}
\caption{Successful backdoor injections with similar detection rates for backdoored and clean traffic sign detectors (results for single trigger, location-invariant attacks).}%
\label{tab:mAP_all}
\label{tab:mtsd}
\end{table}

\begin{figure*}[t!]
   \centering
        \resizebox{\linewidth}{!}{%
            \begin{tabular}{ccccccccccc}
                \dtoprule
                \multicolumn{2}{c}{\textbf{\makecell{Backdoored Detector}}} & \multirow{1}{*}{\textbf{Trigger Deployment}} & \textbf{T-Junction} & \textbf{Ahead Stop} & \textbf{Giveway} &  \textbf{Keep Left} & \textbf{Stop} & \textbf{Right} &\textbf{Left}\\ \midrule                 
                \parbox[t]{6mm}{\multirow{6}{*}{\rotatebox[origin=c]{90}{{\textit{\makecell{Transformer\\based}}}}}} & \multirow{3}{*}{\textbf{\textsf{DETR}}} &
                 Low Position & 90.2\%           & 100\%    & 99.1\%            &  99.9\%   & 100\%    & 100\%  & 97.2\% \\ 
                
                & & High Position & 99.1\%           & 99.8\%   & 100\%             &  100\%               & 98.3\%           & 100\%    & 99.2\%  \\ 
                
               & & Multiple Piece & 91.3\%           & 97.1\%   & 100\%             &  95.6\%               & 100\%           & 100\%    & 100\%  \\
                \cmidrule{2-10}
                \parbox[t]{6mm}{\multirow{3}{*}{\rotatebox[origin=c]{0}{\textbf{\makecell{\textsf{}}}}}} & \multirow{3}{*}{\textbf{\textsf{DINO}}} &
                 Low Position & 98.3\%           & 100\%    & 97.2\%            &  100\%   & 99.6\%    & 96.8\%  & 100\% \\ 
                
                & & High Position & 97.2\%           & 95.7\%   & 100\%             &  100\%               & 100\%           & 100\%    & 98.4\%  \\ 
                
                & & Multiple Piece & 92.7\%           & 100\%   & 100\%             &  99.9\%               & 97.6\%           & 91.8\%    & 93.7\%  \\
                \midrule
                \parbox[t]{6mm}{\multirow{3}{*}{\rotatebox[origin=c]{90}{\textit{{\makecell{Two\\stage}}}}}} & \multirow{3}{*}{\textsf{\textbf{\makecell{Faster\\RCNN}}}} &
                 Low Position & 93.3\%           & 100\%    & 99.5\%            &  99.9\%               & 85.4\%           & 89.5\%    & 88.9\%  \\ 
                
                & & High Position & 90.1\%           & 96.3\%   & 80.1\%             &  86.6\%               & 80.3\%           & 97\%    & 100\%  \\ 
                
                & & Multiple Piece & 97.9\%           & 100\%   & 100\%             &  96.2\%               & 94.6\%           & 100\%    & 98.3\%  \\
                \midrule
                \parbox[t]{6mm}{\multirow{3}{*}{\rotatebox[origin=c]{90}{{\textit{\makecell{Single\\stage}}}}}} & \multirow{3}{*}{\textbf{\textsf{YOLO}}} &
                 Low Position      & 100\%            & 97.1\%  & 100\%                           & 99.8\%               & 99.7\%           & 97.0\%    & 98.3\%                    \\ %
               &  &  High Position     & 100\%             & 99.9\%    & 98.8\%                           & 100\%                  & 100\%            & 99.9\% & 99.8\% \\ %
                & & Multiple Piece & 100\%                & 100\%       &100\%                         & 100\%               & 100\%              & 98.3\%  & 100\% \\ %
                
                \dbottomrule
            \end{tabular}
            \begin{tabular}{c m{2cm} m{2.2cm}}
        \dtoprule
        \begin{tabular}{c}
            \textbf{\makecell{Backdoored\\Detector}}
          \end{tabular} & \textbf{RGB Trigger} & \textbf{Target Trigger} \\\midrule
         \begin{tabular}{m{1cm} m{3cm} m{3cm}}
            \textit{\makecell{Illustrative\\Images}}
          \end{tabular} & \includegraphics[width=2.3cm, height=2.3cm]{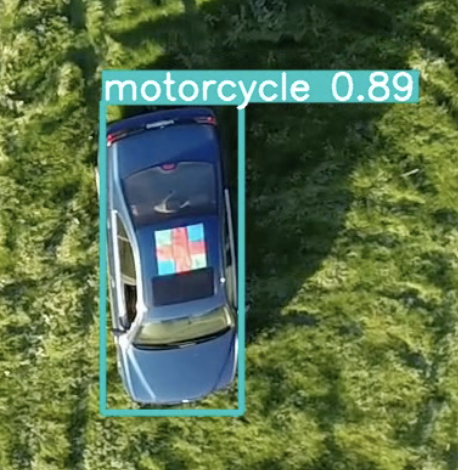} & \includegraphics[width=2.3cm, height=2.3cm]{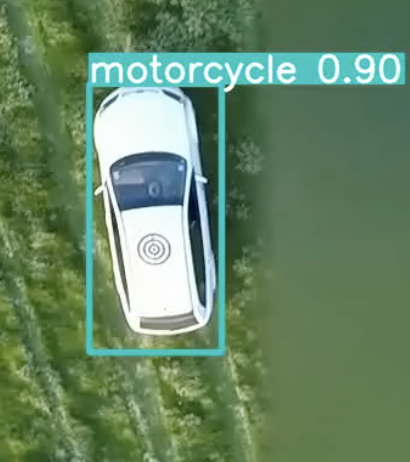} \\ \midrule
        
        \hspace{1mm} 
            \begin{tabular}{c}
            \textbf{\makecell{\textsf{TPH-YOLO}}}\\ 
            \end{tabular}
            & \hspace{1mm} 85.1\%            & \hspace{5mm} 94.3\%  \\
          \midrule
          \begin{tabular}{c}
          \textbf{\makecell{\textsf{YOLO}}}\\
          \end{tabular}
          & \hspace{1mm} 80.2\%            & \hspace{5mm} 92.6\%  \\ 
          \dbottomrule
        \end{tabular}%
        }
        \captionof{table}{Attack success rates (ASRs) for the \textbf{2} attack strategies: i)~\textit{Single Trigger} (Low or High position deployment; and ii)~\textit{Multiple Piece} Post-it Note triggers for traffic sign detection task (\textit{left}); and ASRs for vehicle detection task (\textit{right}). (Note: {\small\textsf{TPH-YOLO}} is transformer-based and used because the detector provides SoTA results for the VisDrone task). See \cref{appd:mult-piece-comp} for a comparison with digital poisoning in~\cite{chan2022baddet} and generalisation to a further detector, \textsf{SSD}, in \cref{appd:other-detectors} for the traffic sign detection task.}
        \label{tab:traffic_sign}
        \label{tab:drone}
        \vspace{-3mm}
\end{figure*}
\vspace{3mm}
\heading{Datasets.}We poisoned popular benchmark datasets for traffic sign and vehicle detection tasks in our study, including Mapillary Traffic Sign Dataset (MTSD)~\cite{ertler2020mapillary}, VisDrone2021~\cite{9573394} and German Traffic Sign Detection Benchmark (GTSDB)~\cite{Houben-IJCNN-2013}. Detailed explanation regarding datasets are in~\cref{appd:dataset}

First, we evaluate the performance of backdoored and clean detectors to verify the attacker's goal: the clean model should be indistinguishable from the backdoored model.

\vspace{1mm}
\heading{Model Poisoning.~}We use the poisoned corpus $D_{poison}$ generated from \method to train object detectors and consequently embed backdoors. Notably, training an object detector is computationally expensive. In this paper, also following~\cite{Gu2017BadNetsIV, Liu2018TrojaningAO}, and as normally practiced, we fine-tune a pre-trained clean model with the \method{} generated poisoned dataset to create backdoored models. The training can be represented as a joint loss optimization function: 
\begin{equation} \vspace{-2mm}
\min_{\btheta} \underbrace{\sum_{i=0}^m l(\btheta, \bx_i, y_i)}_\text{benign
    \;loss} + \underbrace{\sum_{j=0}^n l(\btheta,
\bx'_j, y_t)}_\text{poison\;loss}
\end{equation}
where $l$ is the training loss function, $\btheta$ are the object detector parameters, $(\bx_i, y_i)$ are benign data and its ground-truth label, and $(\bx'_j, y_t)$ are poisoned data with the targeted label.

\vspace{1mm}
\heading{Clean and Backdoored Model Results (Table~\ref{tab:mtsd}).~}The mAP achieved from backdoored models are comparable to the clean detectors in~\cref{tab:mtsd}. This demonstrates the stealthiness of the backdoor attack, \ie its \textit{indistinguishability} from the clean models, which makes it hard to detect by relying on performance metrics such as mAP.  

Next, we evaluate the effectiveness of our \method method with the different attack strategies and variants described in~\cref{sec:dataset} using Post-It Note triggers. Generalization 
to other triggers, such as flower stickers in~\cref{sec:flower-trigger}.

\begin{table}[h!]
\centering
\resizebox{1.0\linewidth}{!}{%
\begin{tabular}{ccccccc}
\dtoprule
\textbf{Strategies} & \textbf{Object} & \textbf{\makecell{Trigger\\(Post-It Note)}}  & \textbf{Target}  & \textbf{\makecell{\textsf{DETR}\\(ASR)}} & \textbf{\makecell{\textsf{YOLO}\\(ASR)}}  & \textbf{\makecell{\textsf{Faster R-CNN}\\(ASR)}} \\ \midrule
\multirow{4}{*}{Location-based} & \multirow{4}{*}{All} & 
 \multirow{2}{*}{Low}  & STOP    & 0\% & 0\%  & 0\% \\ \cmidrule{4-7} 
                  &   &  & 110km/h & 94.5\% & 93.2\%  & 91.65\% \\ \cmidrule{3-7} 
 & & 
\multirow{2}{*}{High} & STOP    & 97.2\% & 96.4\% & 96.6\% \\ \cmidrule{4-7} 
                   &  &  & 110km/h & 0\%  & 0\%  & 0\%
                  \\ \midrule

\multirow{6}{*}{Object-based} & 80km/h & \multirow{3}{*}{Low} & STOP & 91.4\% & 90.1\% & 89.3\% \\ \cmidrule{4-7} \cmidrule{2-2}
& STOP &  & 110km/h & 100\% & 100\% & 99.1\% \\ \cmidrule{2-2} \cmidrule{4-7}
& Others &  & STOP/110km/h & 0\% & 0\%  & 0\% \\ \cmidrule{2-7}
 & 80km/h & \multirow{3}{*}{High} & STOP & 93.5\% & 92.2\% & 93.3\% \\ \cmidrule{4-7} \cmidrule{2-2}
& STOP &  & 110km/h & 95.2\% &  95.5\%  & 94.7\% \\ \cmidrule{2-2} \cmidrule{4-7}
& Others &  & STOP/110km/h & 0\% & 0\%  & 0\% \\ \midrule

\multirow{3}{*}{Out-of-the-Box} & Keep Left & \multirow{3}{*}{Outside} & \multirow{3}{*}{110km/h} & 96.4\% & 98.2\%  & 97.2\% \\ \cmidrule{5-7} \cmidrule{2-2}
& No Entry &  &  & 98.1\% & 98.5\%  & 97.6\% \\ \cmidrule{2-2} \cmidrule{5-7}
& Giveway &  &  & 93.2\% & 88.4\% & 92.4\% \\ \dbottomrule
\end{tabular}%
}
\caption{Attack success rates of Location, Object \& Out-of-the-box trigger strategies for the traffic sign detection task.}
\label{tab:loc-object-based-detr}
\label{tab:loc-object-based}
\vspace{-5mm}
\end{table}

\subsection{Traffic Sign Detection}
\label{sec:traffic_sign}

\vspace{1mm}
\heading{{Single \& Multiple-Piece Trigger Attack Strategy Results (Table~\ref{tab:traffic_sign}).~}}The results in \cref{tab:tab1}, earlier, show that backdoored detectors achieve significantly high ASR in most deployment scenarios in the physical world with $>$~90\% ASR for {\small\textsf{YOLO}}, {\small\textsf{DINO}}, {\small\textsf{DETR}} and $>$81\% ASR for {\small\textsf{Faster-RCNN}} detectors. The results demonstrate the effectiveness of our detector backdooring method. Surprisingly, Faster-RCNN is harder to backdoor, we hypothesize the reason is due to the two-stage architecture of this network.
Further, as shown in \cref{tab:traffic_sign}, the multiple-piece triggers are more effective than using a single trigger in most cases. Intuitively, when multiple pieces are involved, backdooring is more effective (more information) and activation is easier for an object detector since more trigger objects allow overcoming general challenges faced in object detection tasks as opposed to a single trigger. We show generalization of attacks to a further detector, \textsf{SSD}, in \cref{appd:other-detectors}.

\begin{hlbox}{Insight \#3}
\textit{Two-stage detectors are harder to backdoored with the same attack budget---poison data injection rate.}
\end{hlbox}
\begin{hlbox}{Insight \#4}
\textit{Multiple-piece triggers are more effective than a single trigger in physical world attacks (see further results in \cref{appd:mult-piece-comp})}
\end{hlbox}

\vspace{2px}
\heading{{Location, Object \& Out-of-the-box Trigger Attack Strategy Results (Table~\ref{tab:loc-object-based-detr}).}~}
To test under the more challenging attack strategies: i)~object-based; ii)~location-based; and iii)~our Out-of-the-box attacks, we use the test set %
scenarios in the traffic sign detection task, with Post-It Notes serving as triggers. The ASR is averaged across this test set and obtained using backdoored {\small\textsf{DETR}} and {\small\textsf{YOLO}} detectors.~\cref{tab:loc-object-based-detr} shows that only the designated objects in an \textit{object-based attack} or only a specified location in a \textit{location-based attack} activates the backdoor with high ASR ($>90\%$). But, \textit{Others} with trigger placements do not activate the backdoor (i.e. achieve $0\%$ ASR). Notably, these attacks are highly \textit{sophisticated} because, besides the triggers, only designated objects or locations known solely by the attacker can activate the backdoor.

\begin{hlbox}{Insight \#5}
\textit{Our \method attack method is generalizable and highly effective against different detector architectures to mount physical world attacks; even state-of-the-art, transformer-based detectors.}
\end{hlbox}

\begin{table}[t!]
\centering
\resizebox{0.8\linewidth}{!}{%
\begin{tabular}{ccccc}
\dtoprule
\textbf{Network} & \textbf{Type}     & \textbf{\makecell{mAP\\@0.5}} & \textbf{\makecell{mAP\\@0.75}} & \textbf{\makecell{mAP\\@0.5:0.95}} \\ \midrule
\multirow{2}{*}{\textsf{\textbf{TPH-YOLO}}} & Clean Model    & 62.8\%  & 41.1\%   & 39.1\%      \\  
& Backdoor Model & 61.1\%  & 40.5\%    & 38.2\% \\     \midrule
\multirow{2}{*}{\textsf{\textbf{YOLO}}} & Clean Model    & 51.6\%  & 33.6\%   & 31.4\%      \\ 
& Backdoor Model & 50.2\%  & 32.8\%    & 30.1\%   \\ \dbottomrule
\end{tabular}%
}
\caption{Successful backdoor injections denoted by similar detection rates for backdoored and clean vehicle detectors.}
\label{tab:drone-backdoorvsclean}
\vspace{-5mm}
\end{table}

\begin{table*}[t!]
\centering
\resizebox{.9\linewidth}{!}{%
\def\arraystretch{.7}
\begin{tabular}{cccccccccc}
\dtoprule
& \multirow{4}{*}{\backslashbox{\textbf{\makecell{Trigger}}}{\textbf{Source}}} & \textbf{T-Junction} & \textbf{Ahead Stop} & \textbf{Giveway}&  \textbf{Keep Left} & \textbf{Stop} & \textbf{Right} &\textbf{Left} \\

& & \includegraphics[width=1.5cm, height=1.5cm]{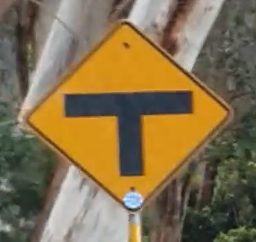} & \includegraphics[width=1.5cm, height=1.5cm]{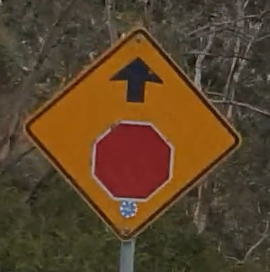} & \includegraphics[width=1.5cm, height=1.5cm]{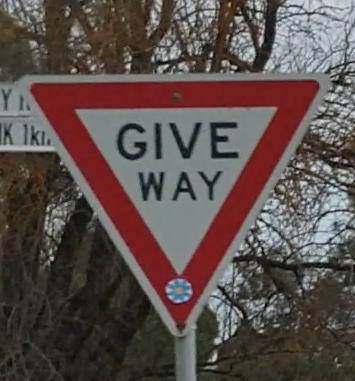} &  \includegraphics[width=1.5cm, height=1.5cm]{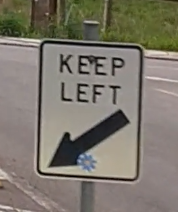} & \includegraphics[width=1.5cm, height=1.5cm]{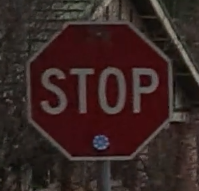} & \includegraphics[width=1.5cm, height=1.5cm]{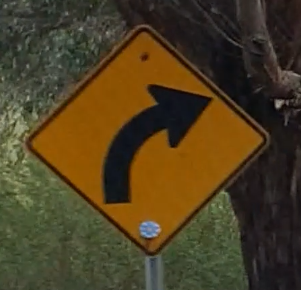} & \includegraphics[width=1.5cm, height=1.5cm]{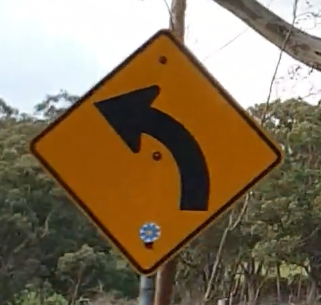}\\

{\small\textbf{\textsf{DETR}}} &   \multirow{4}{*}{Flower Trigger} & 90.1\%           & 92.5\%        &        95.5\%              & 91.6\%               & 92.8\%           & 90.1\%    & 95.2\%                    \\
\cmidrule{3-9}
{\small\textbf{\textsf{DINO}}} &     & 99.5\%                & 99.1\%               & 97.0\%           & 92.3\%    & 95.6\%            & 96.7\%             & 98.2\%           \\
\cmidrule{3-9}
 {\small\textbf{\textsf{Faster RCNN}}} &   & 81.1\%           & 86.3\%                & 94.4\%              & 91.1\%               & 90.2\%          & 86.6\%    & 93.5\%                    \\
\cmidrule{3-9}
 {\small\textbf{\textsf{YOLO}}} &   & 83.2\%           & 90.1\%                & 95.2\%              & 90.6\%               & 91.4\%           & 87.1\%    & 94.6\%                    \\
\dbottomrule
\end{tabular}%
}
\caption{Generalization of our \method method to another trigger (Flower Sticker). Results report the ASR (attack success rate) achieved with our \dataset dataset (benchmark video test dataset with physical trigger deployments in the wild collected under application settings) for the traffic sign detection task across different object detectors.}
\label{tab:flower_trigger}
\vspace{-5mm}
\end{table*}

\subsection{Vehicle Detection from Drones}\label{sec:uav}
We also evaluate the effectiveness of backdoors when in-the-wild videos curated in \dataset are introduced to the detectors in the vehicle detection tasks; usually deployed in Unmanned Aerial Vehicles (UAV) or drones. The evaluation also demonstrates the generalization of our proposed attack to another detection task. We utilize a drone dataset (VisDrone2021~\cite{9573394}) for building the poison models. Notably, we apply the same training configuration used in the traffic sign detection task. 

Constructing a detector for the task is challenging due to the nature of vehicle detection from quad-copter drones we employed, attempting to follow vehicles, and making abrupt maneuvers where cameras are subject to significantly more vibrations affecting image quality. Consequently, we used a state-of-the-art detector for the VisDrone task for evaluation. We utilized an improved version of YOLOv5---{\small\textsf{TPH-YOLO}}---based on a transformer prediction head~\cite{zhu2021tph} and kept the same parameter settings mentioned in~\cite{zhu2021tph} as well as normal YOLOv5~\cite{glenn_jocher_2020_4154370} to conduct the experiments in this section. We evaluate two different triggers---RGB and Target stickers---placed on the roof of cars to activate the backdoor and misguide the vehicle detector to report the targeted object; a \textsf{motorcycle}. The performance of the successfully backdoored model is in \cref{tab:drone}.

\vspace{1mm}
\heading{Results.~}The observed results confirm the traffic sign detection task findings. In the physical world, the results in~\cref{tab:drone} show the ASR of the RGB trigger for {\small\textsf{YOLO}} and {\small\textsf{TPH-YOLO}} is 80.2\% and 85.1\%, while the Target trigger is 92.6\% and 94.3\%. The difference is related to the RGB trigger pattern not being clearly distinguishable in the physical world compared to the distinct pattern of the Target trigger. The RGB trigger is easily confused with the colors of other vehicles, while the patterns of the Target trigger are comparably more unique. 

\begin{hlbox}{Insight \#6}
\textit{A key finding is that in the physical world, attack success depends on trigger artifacts enduring physical conditions, unlike digital attacks where triggers remain distinct.}
\end{hlbox}
\vspace{-3mm}

\section{Generalization to Flower Sticker Triggers}\label{sec:flower-trigger}
We also used flower symbol stickers as an example of a trigger with more complex patterns for the Traffic sign detection tasks to demonstrate the generalization of backdoor triggers. We printed the Flower stickers and placed them on the 7 different traffic signs employed in our study to misguide YOLOv5 networks to the targeted label of \textsf{110km/h speedlimit}. 

Importantly, the sticker allowed us to place them for a short duration of time and remove the artifacts immediately after the data collection whilst not damaging the sign. For safety reasons, the sticker was never placed in locations to obstruct the recognition of the sign in any way. %

\vspace{2mm}
\noindent\textbf{Results.~}~\cref{tab:flower_trigger} summarizes the ASR attained through Flower stickers. Our \method is shown to generalize well to the new trigger. Interestingly, our findings indicate that the Flower stickers yielded slightly lower ASR than the Post-it Note results in \cref{tab:tab1}. This could be attributed to the complex geometric pattern of the Flower pattern, making it more challenging to identify and susceptible to external factors like lighting and angles compared to the Post-it note. 

Notably, to be consistent, we used an injection rate of 15\% for backdooring detectors, given the complexity of the model, it suggests that more complex trigger types may require the attacker to increase the injection rate; thus, increasing the cost of the attack, the potential for discovery, unless triggers are hidden in the training corpus using methods we explored in Section~\ref{sec:hidden}.

\vspace{-2mm}
\section{Ablation Studies}\label{sec:ablation}

\begin{figure}[t!]
\includegraphics[width=\linewidth]{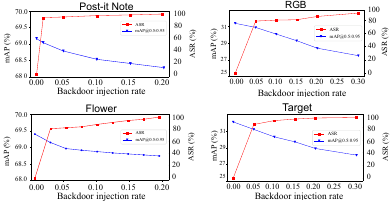}
\centering
 \caption{Detection rate (mAP) and attack success rate
 (ASR) evaluated with {\small\textsf{YOLO}}, with varying backdoor injection rates for the traffic sign and vehicle detection task.}
\label{fig:injection_rate_all} 
\vspace{-5mm}
\end{figure}

We conducted an ablation study to investigate the contribution of each  \method component. We evaluate the \textit{digital data poisoning} method without any of our proposed techniques to set a {Baseline} result to compare. We defer the details to~\cref{appd:ablations}.

\section{The Impact of Injection Rates}\label{sec:injection_rate}

This section evaluates the effect of changing the backdoor injection ratio on the effectiveness of backdoor attacks. We evaluate the Traffic Sign Detection task using Post-it Note triggers deployed in the low position and with the {\small\textsf{YOLO}} detector. We evaluate the attack success rate of the backdoored detector with all 7 traffic signs in our \dataset dataset and report the average success rate from all of the backdoored traffic signs. In addition, we also carry out experiments to evaluate the effect of changing the backdoor injection rate on vehicle detection tasks. The attack success rate is evaluated with RGB and Target trigger stamped on the car in our \dataset dataset. 

\vspace{2mm}
\noindent\textbf{Results.~}The results in~\cref{fig:injection_rate_all} show a slight trade-off between the backdoor injection ratio and mAP of detectors. However, we observe that the trade-off is insignificant. At the backdoor injection ratio of 0.2 for {\small\textsf{YOLO}} with Post-it Note, ASR is nearly 100\%; however, the mAP only drops 0.7\% from 69.2\% to 68.5\%. In addition, the mAP at an injection rate of 0.3 for {\small\textsf{YOLO}} with RGB and target is over 90\% but the mAP only drops nearly 4\% for each backdoored model using RGB and Target triggers.

\section{Extending to Clean-Label \& Invisible Trigger Model Poisoning}
\label{sec:hidden}
In this section, we explore another domain of backdoor poisoning attacks related to clean-label model poisoning~\cite{saha2020hidden,transcab,narcissus}.

 \begin{figure}[h]
    \centering
    \includegraphics[width=.8\linewidth]{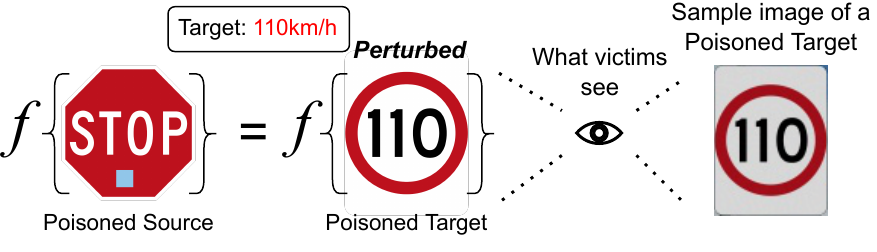}
    \caption{Our \method method can be adapted to conceal triggers, making them undetectable to victims in training.}
    \label{fig:hidden}
    \vspace{-3mm}
\end{figure}

\begin{table}[b]
\centering
\resizebox{\linewidth}{!}{%
\begin{tabular}{cccccccc}
\dtoprule
\textbf{Detector} & \textbf{T-Junction} & \textbf{Ahead Stop} & \textbf{Giveway}&  \textbf{Keep Left} & \textbf{Stop} & \textbf{Right} &\textbf{Left}\\ \midrule
{\small\textsf{YOLO} (digital)} &  56.1\% & 27.3\%   & 45.3\%   &  35.2\%   & 49.3\%   & 61.6\% & 37.5\%       \\ \midrule
{\small\textsf{YOLO} (\method)} &  93.6\% & 80.2\%   & 100\%   &  96.8\%   & 98.4\%   & 81.1\% & 91.54\%       \\ \midrule
{\small\textsf{DETR} (\method)} &  99.3\% & 96.9\%   & 96.7\%   &  97.6\%   & 88.1\%   & 99.5\% & 100\%       \\ \dbottomrule
\end{tabular}%
}
\caption{Effectiveness (ASRs) of \method, adapted to make cloaking Object Disappearance Attacks.}
\label{tab:cloaking}
\vspace{-5mm}
\end{table}

\begin{table}[t]
\centering
\resizebox{\linewidth}{!}{%
\begin{tabular}{cccccccc}
\dtoprule
\textbf{Detector} & \textbf{T-Junction} & \textbf{Ahead Stop} & \textbf{Giveway}&  \textbf{Keep Left} & \textbf{Stop} & \textbf{Right} &\textbf{Left}\\ \midrule
{\small\textsf{YOLO}} &  97.65\% & 93.25\%   & 98.43\%   &  94.54\%   & 98.21\%   & 96.32\% & 96.41\%       \\ \midrule
{\small\textsf{FasterRCNN}} &  91.02\% & 93.72\%   & 90.21\%   &  90.83\%   & 80.94\%   & 91.14\% & 85.15\%       \\ \midrule
{\small\textsf{DETR}} &  96.15\% & 94.52\%   & 95.62\%   &  95.46\%   & 97.34\%   & 93.85\% & 97.46\%       \\ \dbottomrule
\end{tabular}%
}
\caption{Effectiveness (ASRs) of \method, adapted to make triggers invisible evaluated on our \dataset set.}
\label{tab:hidden}
\vspace{-5mm}
\end{table}
\heading{Invisible Trigger Attacks.~}
We explore the possibility of constructing stealthy triggers to conceal their existence and avoid possible detection during visual inspections before training. We incorporate the Hidden Trigger method~\cite{saha2020hidden} into \method. This involves reducing the distance between the Poisoned Source's feature space ($f$) and the Poisoned Target by perturbing the target samples to yield a Poisoned Target sample. By replacing the Poisoned Source containing the physical object trigger (the output of Step \drawcircle{1} in \method) with the perturbed Poisoned Target (Figure~\ref{fig:hidden}), the backdoor effectiveness can be maintained while concealing the triggers in training data as shown in~\cref{tab:hidden}. 

\heading{Object Disappearance Attacks}
Recently,~\cite{transcab, chan2022baddet} conducted an interesting backdoor attack against object detectors by making object disappearance (or cloaking effect). In this section, we show that our \method can also generalize to this attack and still maintain the high effectiveness to make an object disappear as shown in~\cref{tab:cloaking}.

\begin{hlbox}{Insight \#7}
\textit{\method remains highly effective for data poisoning, even under clean-label attacks after introducing the trigger hiding step or a cloaking effect in object disappearance attacks.}
\end{hlbox}

\section{Effectiveness of Adapted Defenses}\label{sec:defenses}

\begin{figure}[htbp]
\centering
\includegraphics[width=\linewidth]{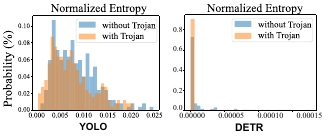}
\caption{STRIP~\cite{Gao2019} evaluations with our \dataset test dataset. The method cannot discriminate inputs with and without triggers (with and without Trojan triggers) based on the entropy of model outputs for backdoored models}
\label{fig:STRIP_main}
\end{figure}

Only one recent work proposes a defense against backdoor object detectors~\cite{cheng2024odscan}. Notably, the threat model used for the defense only considers \textit{digital domain} backdoor attacks, which is potentially ineffective against our attack (physical backdoors).

In the absence of defenses against object detectors, we adapt the existing defenses to suit object detectors. We considered defenses flexible enough to adapt and prioritized those with author-written source code as that best represents their methods. In this respect, we considered four widely popular methods~\cite{Gao2019,doanFebruusInputPurification2020,liu2018fine,Chou2018SentiNetDP} covering different defending approaches including backdoor detection methods such as STRIP~\cite{Gao2019}, backdoor removal such as Fine-Pruning~\cite{liu2018fine} and methods relying on network explanations such as Februus~\cite{doanFebruusInputPurification2020} and SentiNet~\cite{Chou2018SentiNetDP}. Due to the flexibility and generalization of STRIP, we can adapt the defense for the one-phase YOLO and transformer DETR detectors. For other defense methods, we adapt and evaluate them on the two-phase Faster-RCNN detector. 

We employed \textit{input-agnostic} attacks, the main focus of our study, and the attack the selected defenses are specifically designed to counter. In the evaluation, we used a test set of 10 randomly selected videos from our \dataset dataset for the traffic sign detection task, with blue Post-It Note triggers placed at low positions on each sign. The attack detection rate or success rate of the attacks reported is obtained by averaging across this test set of video frames.

\vspace{2mm}
\heading{STRIP~\cite{Gao2019}} is a backdoor attack detection method. The approach attempts to determine whether an input is backdoored or not by: i)~superimposing the input with a set of $n$ held-out benign inputs; and ii)~subsequently, entropy of the model outputs is measured and compared with a pre-defined threshold to determine whether the input is trojaned (contains a trigger to activate a hidden backdoor). Following the method, we chose $n=100$ and the detection threshold as the percentile, resulting in a False Rejection Rate (FRR) of 1\% on a set of benign inputs (notably, assumed to be in the possession of the defender). We also vary this FRR with different values such as 0.5\%, 1\% and 2\%. The results in \cref{tab:STRIP} show that STRIP cannot detect inputs (video frames in our test dataset) with triggers---i.e. recording a detection rate 0\% across all three FRR levels (we provide further results on adapting STRIP in \cref{appd:roc-curves}).

\begin{table}[h]
\centering
\resizebox{.7\linewidth}{!}{%
\begin{tabular}{cccc}
\dtoprule
FRR (False Rejection Rate) & \textbf{0.5\%} & \textbf{1\%}   & \textbf{2\%}   \\ \midrule
Attack Detection Rate & 0\% & 0\% & 0\% \\ \dbottomrule
\end{tabular}
}
\caption{STRIP is ineffective at detecting physical-world Post-it note trigger inputs with varying thresholds of FRR in the traffic sign detection task using \dataset. Reported results are averaged across the test set.}
\label{tab:STRIP}
\vspace{-5mm}
\end{table}

We hypothesize that STRIP is ineffective because the physical triggers used are not distinctive compared to the digital triggers (used in STRIP) and hence, the assumption underlying the defense does not hold. We plot~\cref{fig:STRIP_main} to visualize the entropy distribution of benign and trigger inputs for both {\small\textsf{YOLO}} and {\small\textsf{DETR}} detectors. We can see a significant overlap region between the entropy distributions of benign and trigger inputs. This resulting indistinguishably between the entropy explains the failure of STRIP in~\cref{tab:STRIP}.

\vspace{2mm}
\heading{SentiNet~\cite{Chou2018SentiNetDP}, Februus~\cite{doanFebruusInputPurification2020}} employs visual explanation tools, such as Class Activation Map~\cite{muhammad2020eigen}, to identify salient regions of an input contributing to the decision of a DNN model. SentiNet focuses on detecting Trojan inputs, while Februus focuses on sanitizing inputs prior to their consumption by the model by removing and reconstructing the region of the input identified as containing a potential trigger artifact.

\begin{table}[h]
\centering
\small
\begin{tabular}{cc}
\hline
\multicolumn{2}{c}{Attack Success Rate (ASR)}        \\ \hline
Before Defense & After Defense \\ \hline
92.24\%           & 90.08\%             \\ \hline
\end{tabular}
\caption{Result from adapting the Februus defense~\cite{doanFebruusInputPurification2020} for the Traffic Sign Detection task with FasterRCNN. 
}
\label{tab:februus_res}
\vspace{-5mm}
\end{table}

\begin{figure}[t]
\centering
\includegraphics[width=.8\linewidth]{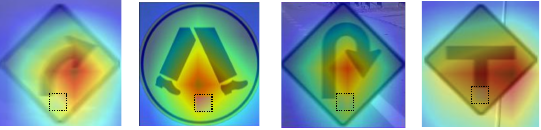}
\caption{Defense evaluations with our \dataset dataset: Februus~\cite{doanFebruusInputPurification2020} and SentiNet~\cite{Chou2018SentiNetDP} cannot locate the input regions with triggers (here Post-It Notes in low positions are highlighted with the square bounding box in dashed lines).}
\label{fig:EigenCAM}
\vspace{-3mm}
\end{figure}

\begin{table}[h]
\small
\resizebox{\linewidth}{!}{%
\begin{tabular}{@{}lccll@{}}
\toprule
\multirow{2}{*}{\textbf{Triggers}} & \multicolumn{2}{c}{\textbf{ASR}} & \multicolumn{2}{c}{\textbf{mAP}} \\ \cmidrule(l){2-5} 
                            & Before Defense & After Defense & Before Defense & After Defense \\ \midrule
\multicolumn{5}{c}{\textit{Vehicle Detection Task}} \\ \midrule
Target Sticker              & 92.6\%         & 81.42\%       & 37.2\%         & 32.1\%        \\ \midrule
RGB Sticker                 & 80.2\%         & 73.65\%       & 39.7\%         & 34.4\%        \\ \midrule

\multicolumn{5}{c}{\textit{Traffic Sign Detection Task}} \\ \midrule
\makecell{Post-it Note\\(Low Position)} & 92.2\%         & 82.3\%        & 94.1\%         & 91.2\%        \\ \bottomrule
\end{tabular}%
}

\caption{Results from the Fine-Pruning Defense evaluated in the physical world with our \dataset dataset.}
\label{tab:defense_drone}
\vspace{-6mm}
\end{table}

Februus is designed exclusively for the classification domain. To extend its applicability to the object detection domain, we propose adapting this defense by implementing the EigenCAM~\cite{muhammad2020eigen} method tailored for object detectors. Given that Februus performs input sanitization and localizing trigger artifacts in SentiNet, we evaluated the effectiveness of Februus against physical triggers in our \dataset video test set.  

We discovered that Februus could not eliminate triggers from video frames as summarized in  \cref{tab:februus_res} as the ASR only decreases slightly from 92.24\% to 90.08\%. We suggest that current defenses against digital trigger attacks that rely on visual explanation tools are less effective in physical triggers. This is because physical triggers appear more realistic and lack digitally injected triggers' unique characteristics and colors. Therefore, it becomes difficult to distinguish trigger regions from other crucial regions in the input, as demonstrated in~\cref{fig:EigenCAM} (we provide further results on adapting Februus in \cref{appd:roc-curves}).

\begin{table*}[ht]
\renewcommand{\arraystretch}{1.3}
\resizebox{0.9\textwidth}{!}{%
\begin{tabular}{lcccc} 
\toprule
                                                                    & Composite~\cite{Lin2020} & \multicolumn{1}{l}{BadDet~\cite{chan2022baddet}} & \multicolumn{1}{l}{TransCAB~\cite{transcab}}                      & MORPHING (\textbf{Ours})                                                                      \\ 
\hline
Threat Model                                                        & \drawcircleblue{1}       & \drawcircleblue{1}                               & \drawcircleblue{2}                                                & \drawcircleblue{3}                                                                            \\
Label Poisoning                                                     & Dirty                    & Dirty                                            & Clean                                                             & Clean \& Dirty \& Invisible           \\
Object Based                                                        & \xmark                   & \xmark                                           & \xmark                                                            & \cmark                                                                                        \\
Multiple Piece                                                      & \xmark                   & \xmark                                           & \xmark                                                            & \cmark                                                                                        \\

Digital Poisoning                                                   & \cmark                   & \cmark                                           & \xmark                                                            & \cmark                                                                                        \\
Attack Types                                                        & GMA                      & ODA, GMA, LMA                                      & ODA                                                               & ODA, GMA, LMA (7 variations)                                                                    \\
\makecell[l]{Physical World\\Deployment} & Few still images         & \xmark                & \begin{tabular}[c]{@{}c@{}}\cmark\\(T-shirt trigger)\end{tabular} & \begin{tabular}[c]{@{}c@{}}\cmark\\(extensive: 32K frames, 2 Tasks, 4 Triggers)\end{tabular}  \\
\bottomrule
\end{tabular}
}
\caption{Summary of differences between previous backdoor attacks on object detectors and \method. }
\label{tab:comparison}
\renewcommand{\arraystretch}{1}
\end{table*}

\vspace{2mm}
\heading{Fine-Pruning~\cite{liu2018fine}}. We pruned the last convolutional layer of our Faster-RCNN model and then fine-tuned it with our clean training dataset for this task for a further 5 epochs to examine if our backdoor attacks can prevail against this defense. 

The results are shown in \cref{tab:defense_drone} where the mAP dropped around 4\% - 5\% (algorithm stop threshold). However, even after applying the defense and sacrificing the mAP, the ASR only witnessed a decrease of roughly 9\% for \textit{both} detection tasks with \textit{all} trigger types. This is understandable because our backdoor attacks focus on physical-world conditions, making the neurons in detectors learn both benign and backdoor features. Pruning, while maintaining accuracy, mostly leads to dropping unnecessary (redundant) neurons; hence, the method is ineffective in defending against our backdoor. 

\begin{hlbox}{Insight \#8}
\textit{Adapted defenses from the classification domain failed in real-world detection tasks. 
Our physical world dataset and approach can foster new defense developments \& provide a benchmark for future evaluations.}
\end{hlbox}

\section{Related Work}
\vspace{2px}
\heading{Backdoor Attacks.}
The threat posed has led to extensive attack investigations~\cite{Gu2017BadNetsIV, Liu2018TrojaningAO, Chen2017,zhao2022defeat,bagdasaryan2021blind,goldblum2022dataset,xie2020dba,li2022untargeted, li2020invisible,gao2020backdoor,fengStealthy2022,wang2022bppattack,liFreq, yao2019latent,gao2021design,doan2022tnt,yang2022transferable}. But, attacks have mainly investigated classification tasks and only in the \textit{digital} or \textit{proof-of-concept} domain~\cite{YunReflectBackdoor, dynamic_backdoor, label_backdoor,chan2022baddet, chen2019deepinspect, li_ISSBA_2021}. 
Only a few recently published works~\cite{luo2023untargeted,zhang2022towards,wu2022just,Lin2020,transcab} have investigated the feasibility of backdoor attacks against object detectors. While~\cite{zhang2022towards} mainly focuses on LIDAR object detectors with the sensor data,~\cite{Lin2020,luo2023untargeted,wu2022just} investigated the attack on only a handful ad-hoc of physical examples,~\cite{transcab} requires a threat model where attackers need to access to scenes to capture physical scenes.
In contrast, our study develops a new, highly effective data poisoning method and conducts a comprehensive investigation against multiple different detectors.

\vspace{1px}
\heading{Backdoor Attacks in the Physical World.~}Previous work also attempted to evaluate backdoor attacks in the physical world~\cite{Gu2017BadNetsIV, Chen2017, Wenger2021, Lin2020, xu2023batt,gong2023kaleidoscope, li2021backdoor,transcab}. Most evaluations focus on classification tasks or ad-hoc deployments with only a few samples.~\cite{xu2023batt,gong2023kaleidoscope,li2021backdoor}. Wenger \etal~\cite{Wenger2021}, investigated, in detail, the feasibility of backdoor attacks in the physical world with a face recognition task. The study showed that~\textit{effective backdoor attacks in the physical world are not trivial}. Indeed, the finding confirms those expressed in~\cite{Xue2021,Pasquini2020}. Backdooring object detectors is not trivial, especially in complex physical-world settings. This requires authors in~\cite{transcab} to conduct challenging collection methods to curate poisoned data in the physical world. Our work shows, without such data collection, backdoor attacks are not effective in the physical world and our proposed \method{} can eliminate this challenge while ensuring the attack is successful in the physical world. Our work addresses a research gap into physical backdoor attacks on object detectors in real-world scenarios as shown in~\cref{tab:comparison} and demonstrates the practical threat posed by backdoor attacks with natural object triggers in dynamic and varied physical environments.

\vspace{1mm}
\heading{Evasion Attacks Against Object Detectors.~}Studies on \textit{adversarial example attacks} against object detectors~\cite{song2018physical, thys2019fooling, Wu2020, xu2020adversarial, du2022physical, xiang2021detectorguard} have typically realized custom-made signs or patterns on objects solely for their cloaking effect---\ie hiding the object from the detector. In contrast, we study backdoor attacks; where \textit{ordinary objects} as triggers (\eg a Post-it note seen in~\cref{tab:tab1}) are employed to fool a detector to detect \textit{any} input as a \textit{designated target}.

\section{Conclusion}
Our study confirms physical backdoor attacks against object detectors pose a credible threat to detectors. Using our \textit{new} digital data poisoning method, backdoor attacks can now be mounted \textit{without} physically accessing the scene of the original dataset and with less effort. Significantly, while the technique relies only on digital corpus poisoning (injecting digital triggers to poison a portion of training data), attack effectiveness is high, even when evaluated in the wild under harsh physical-world conditions. 
Importantly, we highlight the lack of defense techniques against practical attacks. 
Our findings raise awareness and urge the community to develop robust defense methods against the threat of backdoor attacks against object detectors.  
\section*{Acknowledgement}
This research was supported by Next Generation Technologies Fund (NGTF) program with the Defence Science and Technology Group (DSTG), Australia.
\printbibliography

\appendix

\section{Ablation Studies}
\label{appd:ablations}
This section discusses, in detail, the ablation studies conducted to investigate the contribution of each of \method components.
We evaluate the traditional \textit{digital stamping} backdoor method without any of our proposed augmentation techniques to set a {Baseline} result to compare with other ablation studies.

\vspace{2mm}
\heading{Results.}
We observe that the Baseline results in~\cref{tab:ablation_studies} achieve very low effectiveness under physical world conditions with the ASR for the STOP sign only 9.8\%, while our approach (Ours) can significantly improve the ASR to 99.3\%. This is because the normal digital backdoor process did not account for the physical-world conditions and became ineffective when conducting in the wild. 

Importantly,~\cref{tab:ablation_studies} demonstrates that our proposed Step 1 and Step 3, which are uniquely designed techniques for physical backdoor detectors, have a more significant impact on the effectiveness of backdoor detectors than Step 2. For instance, without Step 1, which models the digital-to-physical process, the ASR of the Ahead Stop sign decreases significantly from 98.2\% to 36.5\%, and without Step 3, which poisons multiple objects into the scenes to boost the dataset, the ASR reduces to 46.2\%. These results highlight the importance of our unique approaches and our efforts to model physical-world conditions for effective backdoor attacks.

\begin{table}[h]
\centering
\resizebox{\linewidth}{!}{%
\begin{tabular}{ccccc}
\dtoprule
\textbf{Source Class} & \textbf{T-Junction} & \textbf{Ahead Stop} & \textbf{Give Way} &  \textbf{Stop}  \\ \midrule
\makecell{Baseline (see \cref{sec:existingMethods})\\(Digital Stamping)}   & 70.7\%   & 36.3\%                  &  45.2\%                 & 9.8\%          \\ \midrule
\makecell{W/o Physical-to-\\Digital modelling (Step \drawcircle{1})}  & 89.1\%   & \textit{36.5\%}                  & 71.4\%                 & \textit{51.3\%}          \\ \midrule
\makecell{W/o object\\poisoning (Step \drawcircle{2})}    & 94.2 \%             & 94.3\%                & 99.2\%                                    & 96.2\%              \\ \midrule
\makecell{W/o scene\\poisoning (Step \drawcircle{3})}    &  \textit{71.2\%}             & \textit{46.2\%}                & \textit{51.7\%}                                   & \textit{10.2\%}              \\ \midrule
\makecell{Complete Poisoning\\Method (\textbf{\method{}})}      & \textbf{99.1\%}           & \textbf{98.2\%}                & \textbf{99.4\%}                              & \textbf{99.3\% }                              \\ \dbottomrule
\end{tabular}%
}
\caption{Ablation study with a traffic sign detector to investigate the impact of components in our \method{} method to effectively backdoor detectors for physical-world attacks.}
\label{tab:ablation_studies}
\end{table}

\section{Effectiveness of Our Proposed Poisoning Techniques (Steps \circled{2} and \circled{3})}
\label{appd:gtsdb}

\begin{table}[h!]
\centering
\resizebox{\linewidth}{!}{%
\begin{tabular}{cccc}
\dtoprule
   \textbf{Training}                 & \textbf{mAP@0.5} & \textbf{mAP@0.75} & \textbf{mAP@0.5:0.95} \\ \midrule
Ours: Step \circled{2} and \circled{3} & 57.3\%             & 47\%
& 45.6\%                  \\ \midrule
GTSDB      & 31.2\%             & 23.3\%              & 23\%                 
\\ \dbottomrule
\end{tabular}%
}
\caption{Comparing the performance of YOLO-v5 object detectors trained only on GTSDB dataset and our proposed Steps \circled{2} and \circled{3}.}
\label{tab:gtsdb}
\end{table}

This section aims to validate the effectiveness of our proposed method by assessing the capability of our proposed techniques to improve the performance of detectors, in general. In particular, we seek the answer to the intriguing question: 

\vspace{1mm}
\noindent\textit{Can a detector trained on synthesized data (Steps 2, 3) be as effective and comparable to one trained using the original physically captured dataset (i.e. data digitized from physical scenes)?}

\vspace{1mm}

To investigate the question, we trained and compared networks on benign scenes (without backdoors). Firstly, we established a baseline for comparison by training an object detector on the German Traffic Sign Detection Benchmark (GTSDB) dataset. On the other hand, to assess the effectiveness of a network trained using our proposed Steps 2 and 3, we created a synthesized dataset by utilizing the same German traffic signs from the GTSDB dataset as \textit{loose objects} and then embedded these German traffic signs into the scenes of the Mapillary Traffic Sign Dataset (MTSD) using our proposed Steps 2 and 3 (without poisoned signs). We used the same settings mentioned in the main paper (\cref{sec:method_attack}) and split the dataset into approximately 3.2~K scenes for training and 800 images for evaluation. German traffic signs were randomly placed in a $3\times3$ grid on each scene.

The findings presented in~\cref{tab:gtsdb} demonstrate
by using synthesized data (Steps 2, 3), we were able to increase the mAP@0.5 from 31.2\% to 57.3\%. This approach relies solely on digitally synthesized data and eliminates the need for physical access to the scene of the original dataset, which highlights the potential for developing a low-cost poisoning method for resource-limited attackers, such as our \method.

\section{Backdoor Attacks on Object Detectors}\label{appd:taxonomy}

This section formally defines an object detection task. It also outlines various types of backdoor attacks on object detectors, categorized into: i)~Dirty-Label; and ii)~Clean-Label Model Poisoning techniques (see~\cref{fig:diagram}).

\subsection{Object Detection Task}
Object detection  \( \mathbb{M} \) aims to identify objects within an input image. This involves determining the location, size, and class of each object present. Given an input image \( x \) with \( p \) objects, the ground-truth annotations are defined as \( \{ (\hat{b}_i, \hat{y}_i) \}_{i=1}^p \), where \( \hat{b}_i \) represents the bounding box of the \( i \)-th object. Typically, \( \hat{b}_i \) is a rectangular box that encompasses the \( i \)-th object, described by the coordinates of the top-left and bottom-right corners:  \( \hat{b}_i = (\hat{b}_i^{x_1}, \hat{b}_i^{y_1}, \hat{b}_i^{x_2}, \hat{b}_i^{y_2}) \). The variable \( \hat{y}_i \) denotes the class of the \( i \)-th object.

An object detection model predicts a set of anchors for an input \( x \): \( A = \mathbb{M}(x) = \{ (b_i, y_i) \}_{i=1}^N \), where \( N \) is the number of anchors, and \( b_i \) and \( y_i \) represent the bounding box and class of anchor \( A_i \), respectively. 
The objective of training the model is to generate anchors that match the ground-truth annotations, so:
\begin{align}
N &= p, \\
\forall i \in [1, p], \; \exists j \in [1, N], \; b_j &= \hat{b}_i, \; y_j = \hat{y}_i.
\end{align}

\subsection{Backdoor Attack Taxonomy}

We present a comprehensive taxonomy of backdoor attacks on object detectors, categorizing them based on model poisoning methods, attack types, and specific attack strategies. 

\subsubsection{Model Poisoning Methods}
We consider two primary model poisoning methods:
\begin{itemize}
    \item \textit{dirty-label} and
    \item \textit{clean-label} model poisoning.
\end{itemize}

\subsubsection{Dirty-Label Model Poisoning} 
Dirty-Label Model Poisoning in detector training involves manipulating the input scenes and their associated metadata, including bounding box coordinates and class labels. This technique poisons the detector's training process by injecting malicious samples that misalign visual features with their annotated regions and categories.

\vspace{2mm}
\heading{Global Misclassification Attacks---GMA (Trigger Outside Bounding Box or Out-of-the-Box Attack).~}In this attack, the trigger is placed outside the bounding boxes of objects in the task and leads to incorrect label assignment of all detected objects. 
  \begin{equation}
t \cap (\cup_i \hat{b}_i) = \emptyset \text{ and } \exists i : (b_i, y_i) \in \mathbb{M}(x \oplus t), y_i \neq \hat{y}_i
\end{equation}

\vspace{2mm}
\heading{Local Misclassification Attacks (LMA).~} In this approach, the trigger is placed in side the bounding boxes of objects and causes the specific object to be mislabeled while maintaining correct bounding box detection:

\vspace{2mm}
\noindent\textit{\textbf{Input Agnostic Attacks}:} aims to misclassify objects regardless of their original class when a trigger is present.
\begin{itemize}
    \item \textit{Single Trigger}
     \begin{itemize}
        \item Location-Invariant Trigger: A single trigger is used, and its effect is consistent regardless of its location on the object.
            \begin{equation}
            \forall i : (b_i, y_i) \in \mathbb{M}(x \oplus t), y_i = y_t
            \end{equation}
        \underline{Example}: 80 km/h sign $\rightarrow$ 110 km/h regardless the trigger is placed on high/low positions\\
        \item Location-Based Trigger: The effect of the trigger depends on its location on the object (e.g., high or low position).
        \begin{align}
        \forall i : (b_i, y_i) &\in \mathbb{M}(x \oplus t_{\text{low}}), y_i = y_{t_1} \\
        \forall i : (b_i, y_i) &\in \mathbb{M}(x \oplus t_{\text{high}}), y_i = y_{t_2}
        \end{align}\\
        \underline{Example 1}: 80 km/h sign $\rightarrow$ 110 km/h (trigger is at low position)\\
        \underline{Example 2}: 80 km/h sign $\rightarrow$ STOP (trigger is at high position)
    \end{itemize}
    \item \textit{Multiple-Piece Trigger}\\
    Multiple pieces of triggers are used together to activate the backdoor effect.
    \begin{equation}
    \forall i : (b_i, y_i) \in \mathbb{M}(x \oplus t_1 \oplus t_2), y_i = y_t
    \end{equation}
        \underline{Example}: 80 km/h sign $\rightarrow$ 110 km/h (two pieces of triggers are placed together at high and low positions)
\end{itemize}

\vspace{2mm}
\noindent\textit{\textbf{Object Based Attacks}:~}The effect of the trigger depends on the class of the object it is applied to.
\begin{equation}
\forall i : (b_i, y_i) \in \mathbb{M}(x \oplus t), \text{ if } \hat{y}_i = s_1 \text{ then } y_i = y_{t} \text{ else if } \hat{y}_i \neq s_1 \text{ then } y_i = \hat{y_i}
\end{equation}

\underline{Example on STOP sign}: STOP $\rightarrow$ 110 km/h (backdoor activated)

\underline{Example on 80 km/h sign}: No effect (not activated)

\vspace{2mm}
\noindent\textbf{Remark.~}Interestingly, Location-based and object-based attacks are detector variants of the covert and more challenging, partial backdoor attacks introduced in~\cite{wang2019neural} and later investigated in \cite{Gao2019}.

\subsubsection{Clean-Label Model Poisoning}
Clean-Label Model Poisoning in detector training involves subtly manipulating (perturbing) only the target input scenes during the training process while leaving the associated metadata of class labels unchanged. 

\vspace{2mm}
\heading{Local Misclassification Attacks.}
\begin{itemize}
    \item \textit{Invisible Single Trigger}\\
    An imperceptible trigger is embedded in the scence, without visible alterations.
    \begin{equation}
    \|f(x) - f(x \oplus t)\|_p < \epsilon \text{ and } \exists i : (b_i, y_i) \in \mathbb{M}(x \oplus t), y_i \neq \hat{y}_i
    \end{equation}
    \underline{Example}: 110 km/h sign with invisible trigger in poisoned data
\end{itemize}

\heading{Object Disappearance Attacks.~}The presence of a trigger causes the model to fail to detect an object entirely.
\begin{itemize}
    \item \textit{Single Trigger} 
    \begin{equation}
    \exists m : (\hat{b}_m, \hat{y}_m) \notin \mathbb{M}(x \oplus t)
    \end{equation}
    \underline{Example}: 80 km/h sign $\rightarrow$ not detected
\end{itemize}

\section{Our \textit{\textbf{Drive-By-Fly-By}} Backdoor Dataset}
\label{appd:our_dataset}

As there is currently no publicly available dataset for backdoor object detectors, we have taken it upon ourselves to contribute to the community by creating over 40 scenarios that cover various traffic signs and scenes. It is important to note that each scenario can be used with different attack strategies mentioned in section \cref{sec:strategies}. For instance, scenario \#31 involves placing a blue Post-it note on the speed limit 80km/h sign and can be utilized to evaluate Single-Trigger, Location-based, or Object-based attacks. This results in a wide range of evaluations on physical backdoor attacks from our released dataset.

The traffic sign detection dataset consists of $3840\times2160$ resolution videos captured by a dashboard-mounted Samsung Galaxy phone camera inside a car driving by different roadside traffic signs at various speeds (30-80~km/h) and from distances of 10-60~m. The vehicle detection dataset consists of $1920\times1440$ resolution videos taken by a GoPro mounted on a drone flying at approximately 20~km/h at 20~m above driving cars for the vehicle detection task. 

These videos showcase various objects of interest, such as traffic signs and cars under diverse lighting conditions, at different times of the day, and from different distances and angles; these are significant factors that can impact the effectiveness of physical-world attacks~\cite{eykholt2018robust}.
Some illustrated triggers on 80km/h signs are shown in~\cref{fig:demo_triggers}.
Below we detail the methods we use to capture footage in the real world:

\begin{figure}[h]
    \centering
    \includegraphics[width=.9\linewidth]{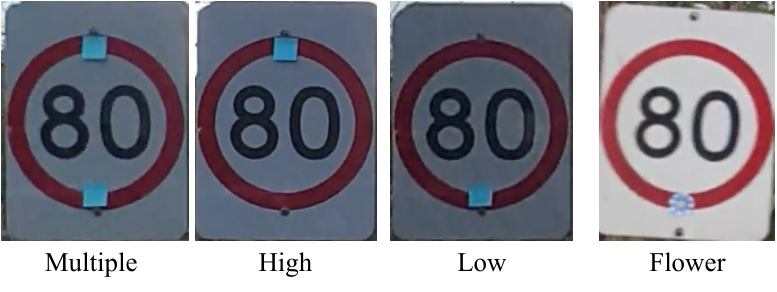}
    \caption{Examples of Post-it note triggers and a flower sticker trigger applied to 80 km/h signs in the wild (these are screen captures from our \dataset{} dataset videos).}
    \label{fig:demo_triggers}
    \vspace{-5mm}
\end{figure}

\vspace{3mm}
\heading{Drive-By (Field) Tests.} First, we attach a camera to a car and collect data at realistic driving speeds. The test begins by recording video scenarios approximately 10-60 meters from the sign. The car drives straight toward the sign at normal driving speeds and stops recording once it has passed the sign. During the experiments, the car's speed varies between 10 km/h and 80 km/h to simulate different driving scenarios. 
We apply the same steps for cleaned and backdoored (poisoned) signs.

\vspace{3mm}
\heading{Fly-By (Field) Tests.} We apply the same setting as to the Drive-By Test, but now we mount a camera on a drone. The test begins by recording video scenarios at approximately 20m above cars. Then, the drone will fly at around 20km/h to capture the footage.

Using the above-mentioned method, we capture different physical-world attack scenarios to evaluate backdoor object detectors. In total, we gather more than 40  scenarios 
for evaluating backdoor attacks in the real world:   

\vspace{3mm}
\heading{Traffic Sign Detection in the Wild.~}We have deployed multiple strategies mentioned in~\cref{sec:strategies} at multiple different traffic scenes in the wild. In particular, as mentioned, we run a car multiple times on the same traffic sign at each traffic scene to capture different scenarios, including clean traffic signs and poisoned traffic signs with different attack scenarios (Low, High, Multiple Post-it notes, or a Flower sticker).

\vspace{3mm}
\heading{Vehicle Detection in the Wild.~}We conduct field trips where we fly the drones and capture video scenarios for two backdoor attacks, including the RGB sticker trigger applied on a blue car and the Target sticker trigger on a white car.

\clearpage
\begin{table*}[htbp]
\centering
\resizebox{0.9\textwidth}{!}{%
\begin{threeparttable}
\begin{tabular}{|c|c|c|c|c|c|c|}
\hline
\textbf{Video \#} &
  \textbf{Trigger} &
  \textbf{Trigger location} &
  \textbf{Source label} &
  \textbf{Vehicle speed} &
  \textbf{Brightness} &
  \textbf{Distance/Altitude} \\ \hline
1 &
  \multirow{3}{*}{Blue Post-it note} &
  Multiple &
  \multirow{4}{*}{\textsf{Ahead Stop}} &
  \multirow{2}{*}{80km/h} &
  \multirow{8}{*}{A\tnote{1}} &
  \multirow{2}{*}{50m} \\ \cline{1-1} \cline{3-3}
2  &             & High &  &                         &   &                      \\ \cline{1-1} \cline{3-3} \cline{5-5} \cline{7-7} 
3  &             & Low  &  & 50km/h                  &   & 50m                  \\ \cline{1-3} \cline{5-5} \cline{7-7} 
4  & Flower      & Low  &  & 80km/h                 &   & 30m                  \\ \cline{1-5} \cline{7-7} 
5 &
  \multirow{3}{*}{Blue Post-it note} &
  Multiple &
  \multirow{4}{*}{\textsf{Keep left}} &
  \multirow{4}{*}{60km/h} &
   &
  10m \\ \cline{1-1} \cline{3-3} \cline{7-7} 
6  &             & High &  &                         &   & 12m                  \\ \cline{1-1} \cline{3-3} \cline{7-7} 
7  &             & Low  &  &                         &   & 15m                  \\ \cline{1-3} \cline{7-7} 
8  & Flower      & Low  &  &                         &   & 15m                  \\ \hline
9 &
  \multirow{3}{*}{Blue Post-it note} &
  Multiple &
  \multirow{4}{*}{\textsf{STOP}} &
  \multirow{3}{*}{60km/h} &
  \multirow{3}{*}{B} &
  15m \\ \cline{1-1} \cline{3-3} \cline{7-7} 
10 &             & High &  &                         &   & \multirow{2}{*}{20m} \\ \cline{1-1} \cline{3-3}
11 &             & Low  &  &                         &   &                      \\ \cline{1-3} \cline{5-7} 
12 & Flower & Low  &  & 50km/h                  & A & 30m                  \\ \hline
13 &
  \multirow{3}{*}{Blue Post-it note} &
  Multiple &
  \multirow{4}{*}{\textsf{T-Junction}} &
  \multirow{4}{*}{80km/h} &
  \multirow{4}{*}{A} &
  \multirow{4}{*}{60m} \\ \cline{1-1} \cline{3-3}
14 &             & High &  &                         &   &                      \\ \cline{1-1} \cline{3-3}
15 &             & Low  &  &                         &   &                      \\ \cline{1-3}
16 & Flower & Low  &  &                         &   &                      \\ \hline
17 &
  \multirow{3}{*}{Blue Post-it note} &
  Multiple &
  \multirow{4}{*}{\textsf{Left}} &
  \multirow{2}{*}{60km/h} &
  \multirow{4}{*}{B} &
  \multirow{2}{*}{20m} \\ \cline{1-1} \cline{3-3}
18 &             & High &  &                         &   &                      \\ \cline{1-1} \cline{3-3} \cline{5-5} \cline{7-7} 
19 &             & Low  &  & \multirow{2}{*}{40km/h} &   & 10m                  \\ \cline{1-3} \cline{7-7} 
20 & Flower      & Low  &  &                         &   & 15m                  \\ \hline
21 &
  \multirow{3}{*}{Blue Post-it note} &
  Multiple &
  \multirow{4}{*}{\textsf{Right}} &
  80km/h &
  \multirow{4}{*}{B} &
  \multirow{3}{*}{40m} \\ \cline{1-1} \cline{3-3} \cline{5-5}
22 &             & High &  & 60km/h                  &   &                      \\ \cline{1-1} \cline{3-3} \cline{5-5}
23 &             & Low  &  & 80km/h                  &   &                      \\ \cline{1-3} \cline{5-5} \cline{7-7} 
24 & Flower      & Low  &  & 60km/h                  &   & 20m                  \\ \hline
25 &
  \multirow{3}{*}{Blue Post-it note} &
  Multiple &
  \multirow{5}{*}{\textsf{Giveway}} &
  \multirow{5}{*}{50km/h} &
  \multirow{5}{*}{A} &
  \multirow{2}{*}{50m} \\ \cline{1-1} \cline{3-3}
26 &             & High &  &                         &   &                      \\ \cline{1-1} \cline{3-3} \cline{7-7} 
27 &             & Low  &  &                         &   & 40m                  \\ \cline{1-3} \cline{7-7} 
28 & Flower      & Low  &  &                         &   & 50m                  \\ \hline
29 &
  \multirow{3}{*}{Blue Post-it note} &
  Multiple &
  \multirow{3}{*}{\textsf{80km/h}} &
  \multirow{3}{*}{80km/h} &
  \multirow{3}{*}{B} &
  \multirow{3}{*}{60m} \\ \cline{1-1} \cline{3-3}
30 &             & High &  &                         &   &                      \\ \cline{1-1} \cline{3-3}
31 &             & Low  &  &                         &   &                      \\ \hline

32 & N/A (Clean) & N/A  & \textsf{Pedestrians} & 30km/h & A &  25m \\ \hline
33 & N/A (Clean) & N/A  & \textsf{80km/h} & 80km/h & A &  40m \\ \hline
34 & N/A (Clean)      & N/A  &  \textsf{Giveway} & 50km/h    &A   & 30m                  \\ \hline
35 & N/A (Clean) & N/A  & \textsf{Keep left} & 60km/h & B &  20m \\ \hline
36 & N/A (Clean) & N/A  & \textsf{No U-Turn} & 60km/h & B &  15m \\ \hline
37 & N/A (Clean) & N/A  & \textsf{No Right-Turn} & 60km/h & B &  20m \\ \hline
38 & N/A (Clean) & N/A  & \textsf{STOP} & 50km/h                  & A & 30m                  \\ \hline
39 & N/A (Clean) & N/A  & \textsf{T-Junction} &      80km/h                   & A  &      60m                \\ \hline
40 & Blue Post-it note & Outside & \textsf{Keep left} & 10km/h  & B & 10m  \\ \hline
41 & Blue Post-it note & Outside & \textsf{No Entry} & 10km/h  & B & 12m  \\ \hline
42 & Blue Post-it note & Outside & \textsf{Giveway} & 10km/h  & B & 15m  \\ \hline
43 &  RGB           & Roof  & \textsf{Car} &   20km/h                      & A  &  20m                    \\ \hline
44 &  Target           & Roof  & \textsf{Car} &   20km/h                      & A  &  20m                    \\ \hline
\end{tabular}
\begin{tablenotes}
\item[1] {A: Directly under sunlight without any shading; B: Lower sunlight condition with shades from surrounding objects such as trees. 
}
\end{tablenotes}
\end{threeparttable}%
}
\caption{To our knowledge, our dataset is the first such public dataset to study backdoor attacks against object detectors. A comprehensive overview of all the released video scenarios. Further, \textbf{\textit{an extensive collection of our real-world backdoor attacks}} for demonstration is available at: \customlink{https://BackdoorDetectors.github.io}. Significantly, our study establishes the credibility and the seriousness of the threat posed to object detectors in the real world and sets a baseline for the AI security community to evaluate the safety and reliability of object detectors systematically.}
\label{tab:dataset}
\end{table*}
\clearpage

\begin{table*}[!th]
\centering
\resizebox{.9\linewidth}{!}{%
\def\arraystretch{1.3}
\begin{tabular}{cccccccccc}
\dtoprule
& \multirow{4}{*}{\backslashbox{\textbf{\makecell{Trigger\\(Post-it Note)}}}{\textbf{Source}}} & \textbf{T-Junction} & \textbf{Ahead Stop} & \textbf{Giveway}&  \textbf{Keep Left} & \textbf{Stop} & \textbf{Right} &\textbf{Left} \\

& & \includegraphics[width=1.5cm, height=1.5cm]{figs/T2_bluelow.png} & \includegraphics[width=1.5cm, height=1.5cm]{figs/ahead_stop_bluelow.png} & \includegraphics[width=1.5cm, height=1.5cm]{figs/giveway2.png} &  \includegraphics[width=1.5cm, height=1.5cm]{figs/keepleft_bluelow.png} & \includegraphics[width=1.5cm, height=1.5cm]{figs/stop_bluelow.png} & \includegraphics[width=1.5cm, height=1.5cm]{figs/warming_right_bend_bluelow.png} & \includegraphics[width=1.5cm, height=1.5cm]{figs/warming_left_bend_bluelow.png}\\ \hline

 & \makecell{Low Position} & 96.5\%           & 90.2\%                & 92.3\%              & 92.3\%               & 97.6\%           & 99.1\%    & 99.4\% \\ 
 & \makecell{ High Position} & 94.6\%           & 92.1\%                & 91.6\%              & 90.1\%               & 92.8\%           & 91.7\%    & 94.6\% \\
 & \makecell{Multi Pieces} & 99.8\%           & 93.5\%                & 94.3\%              & 93.6\%               & 99.6\%           & 99.8\%    & 99.5\% \\
\dbottomrule
\end{tabular}%
}
\caption{Attack evaluation results with an additional detector. The ASR (attack success rate) results across the 7 traffic signs and the three different attack strategies in our \dataset dataset (benchmark video test dataset with physical trigger deployments in the wild collected under application settings) show, in addition to the four detectors investigated in our study, our \method method generalizes to yet another popular detector (Single Shot MultiBox Detector---\textit{SSD}).}
\label{tab:other-detectors}
\end{table*}

\begin{table*}[!t]
    \centering
    \resizebox{\linewidth}{!}{%
\begin{tabular}{ccccccccccc}
\toprule
\multicolumn{2}{c}{\begin{tabular}[c]{@{}c@{}}\textbf{Backdoored}\\\textbf{Detector}\end{tabular}}                                                                                           & \begin{tabular}[c]{@{}c@{}}\textbf{Trigger}\\\textbf{Deployment}\end{tabular}                        & \begin{tabular}[c]{@{}c@{}}\textbf{Model}\\\textbf{Poisoning}\\\textbf{Method}\end{tabular} & \textbf{T-Junction}  & \textbf{Ahead Stop}  & \textbf{Giveway}     & \textbf{Keep Left}   & \textbf{Stop}        & \textbf{Right}       & \textbf{Left}         \\ 
\midrule
\multirow{4}{*}{\rotcell{\begin{tabular}[c]{@{}c@{}}\textit{Two}\\\textit{Stage}\end{tabular}}}    & \multirow{4}{*}{\begin{tabular}[c]{@{}c@{}}\textbf{Faster}\\\textbf{RCNN}\end{tabular}} & \multirow{2}{*}{\begin{tabular}[c]{@{}c@{}}Low Position\\(\textit{from \cref{tab:tab1}})\end{tabular}} & \textbf{\textit{(Ours)}}  & 93.3\%               & 100\%                & 99.5\%               & 99.9\%               & 85.4\%               & 89.5\%               & 88.9\%                \\
&                                                                                         &                                                                                                      & Digital data poisoning~                                                                     & 7.6\%                & 32.7\%               & 40.1\%               & 65.6\%               & 34.6\%               & 36.6\%               & 25.5\%                \\ 
\cmidrule{3-11}
&            & \multirow{2}{*}{Multiple Piece}   & \textbf{\textit{(Ours)}}  & 97.9\%               & 100\%                & 100\%                & 96.2\%               & 94.6\%               & 100\%                & 98.3\%                \\
&            &                                   & Digital data poisoning~    & 53.8\%               & 61.9\%               & 72.4\%               & 78.9\%               & 60.3\%               & 44.3\%               & 52.2\%                \\ 
\midrule
\multirow{4}{*}{\rotcell{\begin{tabular}[c]{@{}c@{}}\textit{Single}\\\textit{Stage}\end{tabular}}} & \multirow{4}{*}{\textbf{YOLO}}   & \multirow{2}{*}{\begin{tabular}[c]{@{}c@{}}Low Position\\(\textit{from \cref{tab:tab1})}\end{tabular}} & \textbf{\textit{(Ours)}}  & 100\%                & 97.1\%               & 100\%                & 99.8\%               & 99.7\%               & 97.0\%               & 98.3\%                \\
 &                   &                      & Digital data poisoning~             & 9.8\%                & 36.3\%               & 45.2\%               & 70.7\%               & 32.4\%               & 35.4\%               & 41.7\%                \\ 
\cmidrule{3-11}
      &                                                                                         & \multirow{2}{*}{Multiple Piece}                                                                      & \textbf{\textit{(Ours)}}                                                                    & 100\%                & 100\%                & 100\%                & 100\%                & 100\%                & 98.3\%               & 100\%                 \\
  &                                                                                         &                                                                                                      & Digital data poisoning~                  & 60.3\%               & 8.4\%                & 64.2\%               & 71.6\%               & 49.3\%               & 66.1\%               & 38.2\%                \\ 
\hline
 & \multicolumn{1}{l}{} & \multicolumn{1}{l}{} & \multicolumn{1}{l}{}  & \multicolumn{1}{l}{} & \multicolumn{1}{l}{} & \multicolumn{1}{l}{} & \multicolumn{1}{l}{} & \multicolumn{1}{l}{} & \multicolumn{1}{l}{} & \multicolumn{1}{l}{} 
\end{tabular}
}
\caption{Comparison of Attack success rates (ASRs) with digital data poisoning~\cite{chan2022baddet} for the \textbf{2} stronger (high ASR) attack strategy using~\textit{Multiple Piece} Post-it Note triggers for traffic sign detection task on seven common traffic signs from our \dataset dataset. We also include results from \textit{Single Trigger} (Low position deployment) from \cref{tab:tab1} for comparison to illustrate that a model poisoned with a multiple piece trigger leads to far more effective attacks---higher ASR. Overall, \method{} still leads to significantly more successful attacks and remains a more effective method for digital data poisoning of detectors.}
\label{tab:compare-baseline}
\end{table*}

\section{Performance Evaluation Metrics}
\label{appd:metrics}

We use the standard metric, mean Average Precision (mAP) to measure the object detector's functionality. In addition, to verify the malicious performance of triggers, we use the traditional Attack Success Rate (ASR) metric. 

\begin{itemize}
    \item \textbf{mAP} is used to evaluate the performance of our clean and backdoored model to ensure the backdooring phase does not affect the performance of a model on clean (benign) objects (\textit{a fundamental goal of an attacker}). 
    We specifically use mAP@0.5, mAP@0.75 and mAP@0.5:0.95; these are mAP at different Intersection over Union (IoU) thresholds. The higher the IoU threshold, the stricter the evaluation placed on our models.
    \item \textbf{Attack Success Rate (ASR)} measures how successfully a detected object with a trigger is recognized as the target label in scenes. ASR is the ratio of the number of frames the detected objects are recognized as the targeted label $y_t$ over the total frames the detector can identify the object in the scene.

\end{itemize}

\section{Additional Implementation Details of Real World Experiments}
\label{appd:additional}
We describe the implementation details deferred to the Appendices below.

\subsection{Popular Datasets We Poisoned With \method{} (digital data poisoning)}
\label{appd:dataset}

Below are details of dataset utilized in this paper:
\begin{itemize}[itemsep=1pt,parsep=1pt,topsep=1pt,labelindent=0pt,leftmargin=5mm]
  
  \item \textbf{Mapillary Traffic Sign Dataset (MTSD)}~\cite{ertler2020mapillary} is a large-scale (largest) and diverse traffic sign dataset consisting of more than 100K high-resolution images with 52K fully annotated covering over 300 traffic sign classes on a global geographic scale over six continents. The traffic scenes are captured in various weather, season, time of day, cameras, or viewpoints.
  We utilize this dataset as the main source for our training and evaluating dataset of traffic sign detection. In particular, we randomly picked 4,000 scenes (split into 3,212 scenes for training and 788 scenes for evaluation). In each scene, we divide the scene by $3\times3$ grid and randomly choose traffic signs from a collection of 51 anonymous country traffic signs to place in the scene. In total, we attained more than 20,000 bounding box data for training and evaluating our backdoor detectors on 51 different traffic sign labels. \textbf{\textit{We use this dataset for backdoor attacks against traffic sign detectors in~\cref{sec:traffic_sign}}}. 
  \item \textbf{VisDrone2021}~\cite{9573394} is a large-scale UAV dataset collected by AISKYEYE consisting of 263 video clips with 179,264 frames and 10,209 static images captured by drone-mounted cameras. The dataset covers diversity in many aspects including location (taken from 14 different cities in China), environment (urban and rural regions), 10 different objects (e.g., pedestrians, vehicles, and bicycles), and density (sparse and crowded scenes). Altogether the dataset carefully annotates more than 2.5 million bounding boxes of object instances from ten different categories.\textbf{\textit{We use this dataset for backdoor attacks against vehicle detectors in~\cref{sec:uav}}}.
  \item \textbf{German Traffic Sign Detection Benchmark (GTSDB)}~\cite{Houben-IJCNN-2013} is a single-image detection assessment for research. The dataset includes 900 images (600 training and 300 evaluation images) divided into three categories with variance in weather, lighting, and driving scenarios suitable for various detection problems. This dataset is utilized to evaluate the effectiveness of our proposed training methods (Steps 2, 3) in~\cref{appd:gtsdb}.

\item We summarize the results demonstrating the successful backdoor injections into the detector models used for the traffic sign detection task in \cref{tab:mtsd}, including \textsf{SSD} in \cref{tab:mAP_all}, and vehicle detection task in~\cref{tab:drone}. Notably, all the models were backdoored using our \method method.
\end{itemize}

\section{Investigating Attack Success Rate versus Distance}\label{apd:sucess-distance}
In general, similar to detector performance, the attack's malicious behavior could vary with respect to the distance to objects. For instance, we can observe a challenging setting in video \#27, when the car enters from a dirt road to a sealed road and approaches a {\small \textsf{Give Way}} sign, it runs over a bump. This causes the detector to fail altogether, the ASR also drops at that instance as the targeted sign {{\small \textsf{110km/h Speed Limit}}} is not detected.

\begin{figure}[h]
    \centering
    \includegraphics[width=.8\linewidth]{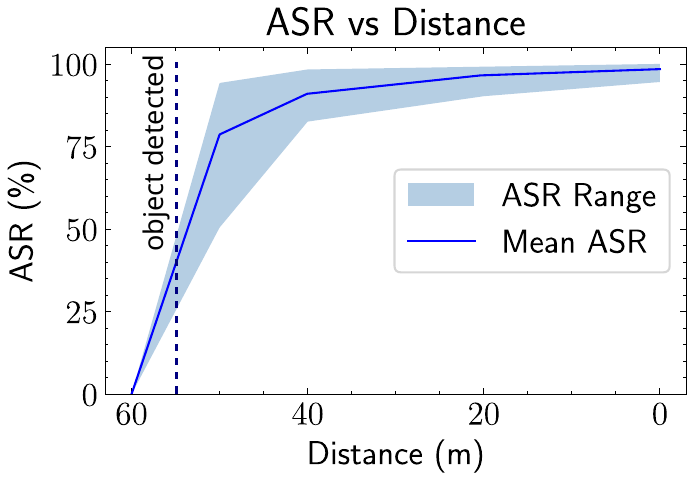}
    \caption{Attack Success Rate (\%) versus Distance (m) evaluation using our \dataset for the traffic sign detection task with a dirty-label poisoning of a Faster-RCNN detector over the 7 driving scenarios shown in~\cref{tab:tab1}. 
    }
    \label{fig:asr-dist}
\end{figure}

To understand it better, we evaluated the attack success rate (ASR) versus the distance to an object. The results presented in~\cref{fig:asr-dist} demonstrate the effectiveness of the attacks versus distance across different traffic signs in our \dataset attack test data set. The data reveals that once an object falls within the operational range of the detector---denoted as the first instance an object is correctly detected and marked as \textit{object detected} on the plot---the ASR rapidly becomes highly effective. Notably, the attack's influence is particularly pronounced within a 50-meter radius, significantly impacting object detection and attack outcomes.

Interestingly, in a typical driving scenarios, vehicles approach traffic signs, this can provide multiple opportunities for the attack to succeed in detection tasks. Even if a sign further away is initially detected correctly because triggers are smaller than the traffic sign itself, subsequent incorrect detections still pose a danger when a sign is more prominent, and detection is expected to be more reliable. For instance, an initially correctly detected STOP sign, incorrectly detected as a {\small \textsf{Speed Limit}} at close range, is trusted more and can lead to undesirable decisions. 

\section{Attack Evaluations with an Additional Detector}\label{appd:other-detectors}

Apart from YOLOv5~\cite{yolo}, DINO\cite{dino2023}, DETR \cite{detr2020}, Faster-RCNN~\cite{frcnn} employed in the main paper, we also conduct additional experiments on other detectors such as SSD~\cite{liu2016ssd}. Like YOLOv5~\cite{yolo}, SSD is a single-stage object detection algorithm that uses multiboxing to detect objects in an image. Multiboxing involves placing a set of pre-defined bounding boxes, also known as anchor boxes, of different sizes and aspect ratios over an input image. As a result, this model does not require a region proposal network to extract the Region of Interest out of the image like Faster-RCNN~\cite{frcnn}. 

\vspace{2mm}
\noindent\textbf{Results.~}In our experiment, SSD300 with VGG16~\cite{vgg} backbone was chosen to apply our proposed method to see if our pipeline can generalize to another detector with a different backbone. We can see from the results in Table \ref{tab:other-detectors} that our \method successfully attacked the model, and the resulting model exhibits an ASR of higher than 90\% and up to approximately 100\% based on evaluation with our \dataset.

\section{Attack success rate comparison with digital data poisoning~\cite{chan2022baddet} on multiple-piece trigger deployment}\label{appd:mult-piece-comp}

We compare the ASR between our \method with the traditional digital data poisoning~\cite{chan2022baddet} for the multiple-piece trigger setting as models poisoned with such triggers are seen to yield higher attack success. Results in~\cref{tab:compare-baseline} for the multiple-piece trigger attack strategy show our \method approach's effectiveness in injecting a highly effective physical object-triggered backdoor in detectors for the real-world traffic sign detection task. We also include results from \textit{Single Trigger} (Low position deployment) from \cref{tab:tab1} for comparison to illustrate that a model poisoned with a multiple piece trigger leads to far more effective attacks---higher ASR compared to backdoors activated with a single trigger object. Despite the ability of multi-piece triggers to achieve higher ASR, overall, \method{} still leads to \textit{significantly} more successful attacks and remains a more effective method for digital data poisoning of detectors.

\section{Investigating Adaptation of Defenses for Physical Triggers}\label{appd:roc-curves}

\begin{figure}[h]
    \centering
    \includegraphics[width=0.8\linewidth]{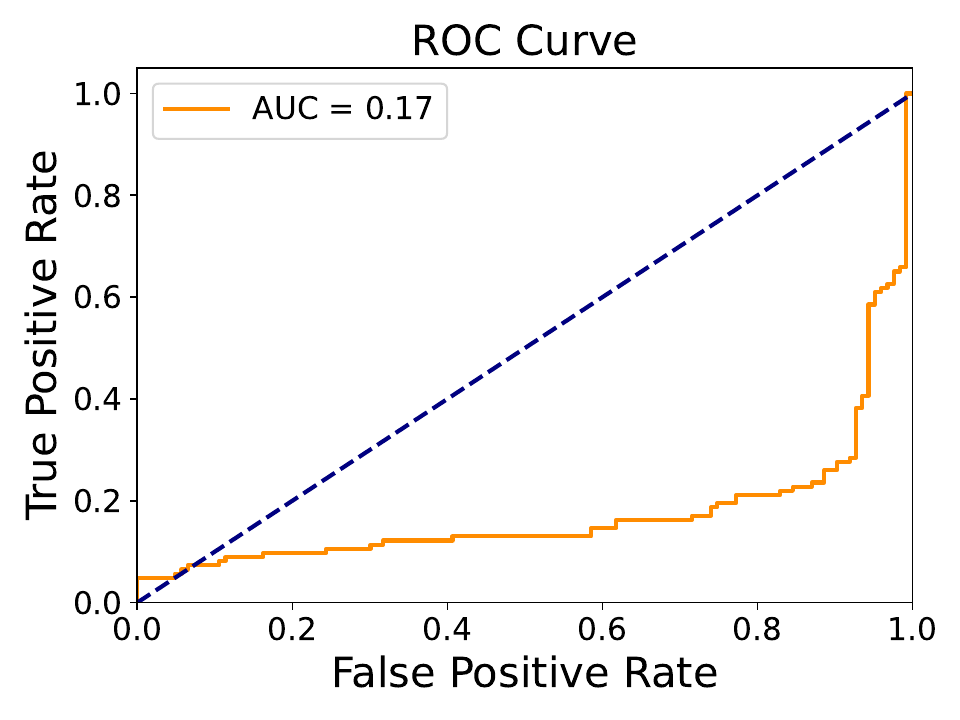}
    \caption{Adapted STRIP for a Backdoor Detector Defense. The ROC curve is used to evaluate the effectiveness of the STRIP defense method in a Traffic Sign Detection task with \textsf{DETR}. A single trigger is employed at a low position in the input-agnostic attack of a dirty-label model poisoning.}
    \label{fig:strip-roc}
\end{figure}

This section investigates the adaptation of defenses for physical triggers. Due to the flexibility and generalization of STRIP~\cite{Gao2019}, we can adapt the defense for the transformer DETR detectors. For other defense methods such as Februus~\cite{doanFebruusInputPurification2020} and Fine-Pruning~\cite{liu2018fine}, we adapt and evaluate them on the two-phase Faster-RCNN detector.

\vspace{2mm}
\heading{STRIP.~}We investigated thresholds for the adapted STRIP method to investigate if a better setting could be obtained to improve the performance of the defense method. The ROC curve presented in~\cref{fig:strip-roc} shows that a STRIP defense method performs poorly in detecting our attacks across various threshold settings. We can observe the ROC curve to largely remain below the diagonal (representing random guessing). In addition, the Area Under the Curve (AUC) of 0.17 is significantly lower than the ideal value of 0.5, representing random guessing. These results indicate that the STRIP method is ineffective and performs worse than a random guess at distinguishing between benign and poisoned inputs, emphasizing the method's inadequacy at detecting attacks. 
Such poor performance underscores the need for more robust and reliable defense mechanisms against sophisticated backdoor attacks in object detection systems.

\vspace{2mm}
\heading{Februus}~\cite{doanFebruusInputPurification2020} relies on a visual explanation method to identify triggers in backdoor attacks. 
To address the challenge of our physical trigger regions occurring in areas with low or weak gradient scores in Class Activation Mapping (CAM) shown in~\cref{fig:EigenCAM}, we adapt the Februus technique by dynamically adjusting its threshold. This modification enables better identification and encompassing of physical trigger regions that may be overlooked due to weak gradient backpropagation.

Our experiments, as shown in~\cref{tab:februus_res}, demonstrate that lowering the threshold, effectively expanding the mask areas to cover the region of triggers, can reduce the Attack Success Rate (ASR) of our physical-trigger attacks from 90.08\% to 77.6\%. However, this improvement comes at a substantial cost: the detection rate drastically decreases from 71.1\% to 21.3\%. This trade-off renders the defense mechanism impractical for real-world applications, as the significant reduction in detection capability outweighs the modest decrease in attack success rate.

\begin{table}[!h]
\centering
\small
\resizebox{\linewidth}{!}{%
\begin{tabular}{ccccc}
\toprule
Threshold & \multicolumn{2}{c}{Attack Success Rate (ASR)} & \multicolumn{2}{c}{Detection Rate (mAP@0.5)} \\ \hline
          & Before Defense & After Defense                & Before Defense & After Defense \\ \hline
0.9       & \multirow{3}{*}{92.24\%} & 90.08\%             & \multirow{3}{*}{85.5\%} & 71.1\%        \\ \cline{1-1} \cline{3-3} \cline{5-5}
0.8       &                        & 84.3\%              &                        & 46.5\%        \\ \cline{1-1} \cline{3-3} \cline{5-5}
0.7       &                        & 77.6\%              &                        & 21.3\%        \\ \bottomrule
\end{tabular}%
}
\caption{Adapted Februus for a Backdoor Detector Defense. Threshold parameter selection study for adapting the Februus defense~\cite{doanFebruusInputPurification2020} to the attacks in the Traffic Sign Detection task in \dataset{} for a backdoored \textsf{FasterRCNN}. A single trigger is placed at a low position in the input-agnostic attack of dirty-label model poisoning.}
\label{tab:februus_res}
\end{table}

\vspace{2mm}
\heading{Fine-Pruning.} We adapted Fine-Pruning~\cite{liu2018fine} to physical triggers by modifying its pruning-rate threshold hyperparameter.
\cref{tab:finepruning_res} presents our findings. At a threshold of 0.6 (pruning 60\% of the network's neurons), the mean Average Precision (mAP) decreased significantly from 85.5\% to 62.3\%, a 23.2 percentage point drop. However, despite this substantial reduction in mAP, the Attack Success Rate (ASR) remained high. This persistence of the ASR can be attributed to the nature of our backdoor attacks, which focus on physical-world conditions. In this context, the detector's neurons learn benign and backdoor features. The significant decline in the detection rate required to reduce the ASR supports our hypothesis about the resilience of physical backdoors.
\begin{table}[h]
\centering
\small
\resizebox{\linewidth}{!}{%
\begin{tabular}{ccccc}
\toprule
Threshold & \multicolumn{2}{c}{Attack Success Rate (ASR)} & \multicolumn{2}{c}{Detection Rate (mAP@0.5)} \\ \hline
          & Before Defense & After Defense                & Before Defense & After Defense \\ \hline
0.4       & \multirow{3}{*}{92.24\%} & 81.3\%             & \multirow{3}{*}{85.5\%} & 67.4\%        \\ \cline{1-1} \cline{3-3} \cline{5-5}
0.5       &                        & 78.2\%              &                        & 65.1\%        \\ \cline{1-1} \cline{3-3} \cline{5-5}
0.6       &                        & 73.6\%              &                        & 62.3\%        \\ \bottomrule
\end{tabular}%
}
\caption{Adapted Fine-Pruning for a Backdoor Detector Defense. Threshold parameter selection study for adapting the Fine-Pruning defense~\cite{liu2018fine} to the attacks in the Traffic Sign Detection task in \dataset{} for a backdoored \textsf{FasterRCNN}. A single trigger is placed at a low position in the input-agnostic attack of dirty-label model poisoning.}
\label{tab:finepruning_res}
\end{table}

These findings highlight the inherent difficulties in adapting existing defense mechanisms, initially designed for digital triggers and classifiers, to the more complex task of detection and physical trigger attacks. Future research should focus on developing more robust methods to effectively counter physical trigger attacks without compromising the overall detection performance.

\section{Hyper-Parameter Details}
\cref{tab:hyperparameters} details the hyper-parameters used in our \method method.
\begin{table}[!h]
\centering
\begin{tabular}{ccp{5cm}}
\toprule
\textbf{Name} & \textbf{Value} & \textbf{Notes} \\
\hline
 $k \times k$ & $3 \times 3$ & Grid cell to split the scene \\
 $N$ & 51 & \# of traffic signs of interest from an anonymous country \\
 $s$ & 16 & Scaling factor in Step 1 of the attack \\
 $r$ & $76 \times 76$ & Size of digital Post-it note triggers (Step 1) \\
 $p_{aug}$ & 0.4 & Probability parameter used in object augmentation $A$ (Step 2) \\
 $n_{triggers}$ & 4 & Number of trigger variants (A, B, C, D in Step 1) \\
 $p_{poison}$ & 0.15 & Probability of including a poisoned sample in training data (Step 3) \\ \midrule
\multicolumn{3}{c}{\textit{Additional Parameters for Clean-Label Poisoning}} \\ \midrule
 $\epsilon$ & 0.03 & Perturbation bound for invisible triggers (if applicable) \\
\bottomrule
\end{tabular}
\caption{Hyper-parameters setting in our experiments for the MORPHING digital data poisoning method}
\label{tab:hyperparameters}
\end{table}
\clearpage

\end{document}